\documentclass{article}

\usepackage{arxiv}

\usepackage[utf8]{inputenc}  
\usepackage[T1]{fontenc}     
\usepackage{hyperref}        
\usepackage{url}             
\usepackage{booktabs}        
\usepackage{amsmath}
\usepackage{amsfonts}        
\usepackage{nicefrac}        
\usepackage{microtype}       
\usepackage{cleveref}        
\usepackage{lipsum}          
\usepackage{graphicx}
\usepackage{natbib}
\usepackage{doi}
\usepackage{listings}
\usepackage{xcolor}
\usepackage{multirow}

\title{Development of low-dissipative projection method framework integrating various high-order time integration schemes using OpenFOAM}

\renewcommand{\headeright}{Research Paper}
\renewcommand{\undertitle}{Research Paper}

\renewcommand{\shorttitle}{%
Low-Dissipative Framework Integrating High-Order Time Schemes}

\author
{%
  \href{https://orcid.org/0000-0001-6232-690}{\includegraphics[scale=0.06]{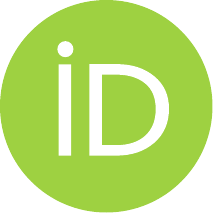}\hspace{1mm}Hao~Guo} \\
	Institute of Engineering Thermophysics \\
	Tsinghua University \\
	Beijing, China \\
	\texttt{guoh19@proton.me} \\
  \AND
  Peixue~Jiang \\
  Institute of Engineering Thermophysics \\
  Tsinghua University \\
	Beijing, China \\
  \texttt{jiangpx@tsinghua.edu.cn} \\
	\And
  Yinhai~Zhu\thanks{Corresponding author.} \\
	Institute of Engineering Thermophysics\\
	Tsinghua University \\
	Beijing, China \\
	\texttt{yinhai.zhu@tsinghua.edu.cn} \\
}

\graphicspath{ {figures/} {./} }

\hypersetup
{%
  colorlinks = true,
  urlcolor   = blue,
  citecolor  = black,
  pdftitle={Development of low-dissipative projection method framework integrating various high-order time integration schemes using OpenFOAM},
  pdfauthor={Hao~Guo, Peixue~Jiang, Yinhai~Zhu},
  pdfkeywords={Low dissipation, High-order time scheme, Linear multi-step method},
}

\begin{document}

\maketitle

\begin{abstract}
	A low-dissipative solution framework integrating various types of high-order time scheme is proposed and implemented based on the open-source C++ library OpenFOAM. This framework aims to introduce different categories of low-dissipative time integration schemes into a unified solver convenient for comparison of scheme performance in finite volume computational fluid dynamics code, contributing to the development of low dissipation scheme appropriate for scale-resolving turbulence simulation. To demonstrate this general framework's ability of including a wide range of time integration method, in addition to typical Runge--Kutta family schemes of linear single-step method, two more complex linear multi-step method, Adams--Bashforth family and Adams--Bashforth--Moutton family schemes, are implemented with the projection algorithm, which increase the options of time discretization. The unified solver obtained by the solution framework select the specified time scheme from a variety of alternatives in runtime rather than maintaining multiple solvers with each compiled for a single scheme, while new scheme can be easily added according to the basic idea of this universal framework. In order to verify framework's effectiveness and practicality, four low-dissipation schemes are adopted to perform a series of test cases. The results show that, compared with the low-order implicit scheme, the corresponding schemes implemented in the universal framework has lower numerical dissipation and more accurate in scale-resolving turbulence simulations. Moreover, among the high-order schemes with the same order of accuracy, the multi-step schemes exhibit a considerable advantage in computation time. Further research on the stability of the explicit scheme indicates that a multi-step method with an appropriate order may be an optimal choice when taking both the prediction accuracy and computational efficiency into account in unstable problems.
\end{abstract}


\section{Introduction}

With the rapid development of high performance computation (HPC) in the past decade, scale-resolving turbulence simulation methods such as detached eddy simulation (DES), large eddy simulation (LES) and direct numerical simulation (DES) have been increasingly applied to more complex flows \citep{Chain:2015,Sun:2019,Guo:2023:HMT}. Because the real engineering problems often encounter complex geometry, computational fluid dynamics (CFD) code based on structured grid is difficult to meet the growing requirements. Therefore, both widely used commercial code (such as ANSYS Fluent \cite{Fluent} and STAR-CCM+ \citep{STAR}) and popular open source code (such as OpenFOAM \citep{FOAM} and code\_saturne \citep{Saturne}) choose to develop solvers based on unstructured collocated grids with higher computation efficiency and lower memory/storage requirements to overcome this limitation.

Previous researches have shown that high-resolution LES and DNS not only have a requirement for computational efficiency, but also demand accurate numerical methods \citep{Kim:1987,Canuto:2007,Jasak:2007,Vuorinen:2014} where high-order time/spatial discretization are particularly important to reduce the influence of numerical dissipation and numerical dispersion on the calculation results \citep{Kim:1987,Komen:2021,Shu:1988:1,Vuorinen:2012}. However, in contrast to the common adoption of high-order scheme in academic research/code, most of the widely used CFD codes for general purpose (e.g., ANSYS Fluent, OpenFOAM) are still only embedded with regular second order method. Such a move may be motivated by stability concerns. However, in order to carry out high resolution LES computation in the industrial applications, many researchers have tried to implement high-order time integration method based on the C++ open source libraries OpenFOAM and have developed corresponding solvers to reduce dissipation \citep{Vuorinen:2014,Kazemi-Kamyab:2015,Alessandro:2018}.

Resent researches on the implementation of high-order time discretization scheme are mainly based on the Runge--Kutta family of linear single-step method. \citet{Vuorinen:2014} combined with the projection algorithm (fraction step method), implemented the classical fourth-order Runge--Kutta scheme and the third-order accelerated Runge--Kutta scheme based on OpenFOAM framework. \citet{Kazemi-Kamyab:2015} implemented implicit Runge--Kutta family schemes with higher order combined with the PISO algorithm \citep{Issa:1986:1}. Then based on both approach of \citet{Kazemi-Kamyab:2015} and approach of \citet{Sanderse:2012}, \citet{Alessandro:2018} developed solvers for implicit and explicit Runge--Kutta schemes for heat transfer problems in incompressible flows. \citep{Castaño:2019} implemented a version of the incremental-pressure RK projection method and compared the performance with the pisoFoam solver of standard OpenFOAM release. The main limitations of existing implementations lie in two points. First, almost all the work only considers the Runge--Kutta family of single-step schemes, and there is a lack of research on other types of high-order schemes, especially the linear multi-step method. Second, unlike the native OpenFOAM solvers which decide the time scheme in the run time based on user's setting, most developed solvers can only use a specific scheme. If a different and/or less dissipative scheme needs to be adopted, a new solver must be rewritten and recompiled (as versions change, there may also be problems with earlier solver ideas not working with the new library). This makes it more difficult to compare the performance of different types of low-dissipation time integration schemes. \citep{Komen:2020} developed a single unified solver, RKFoam, that integrates the schemes of \citet{Vuorinen:2014}, \citet{Kazemi-Kamyab:2015}, and \citet{Alessandro:2018}. But, indicated by the solver's name, the schemes for their unified solver is still based solely on the Runge--Kutta family method.

\begin{figure}
	\centering
  \fbox{
    \includegraphics[width=12cm]{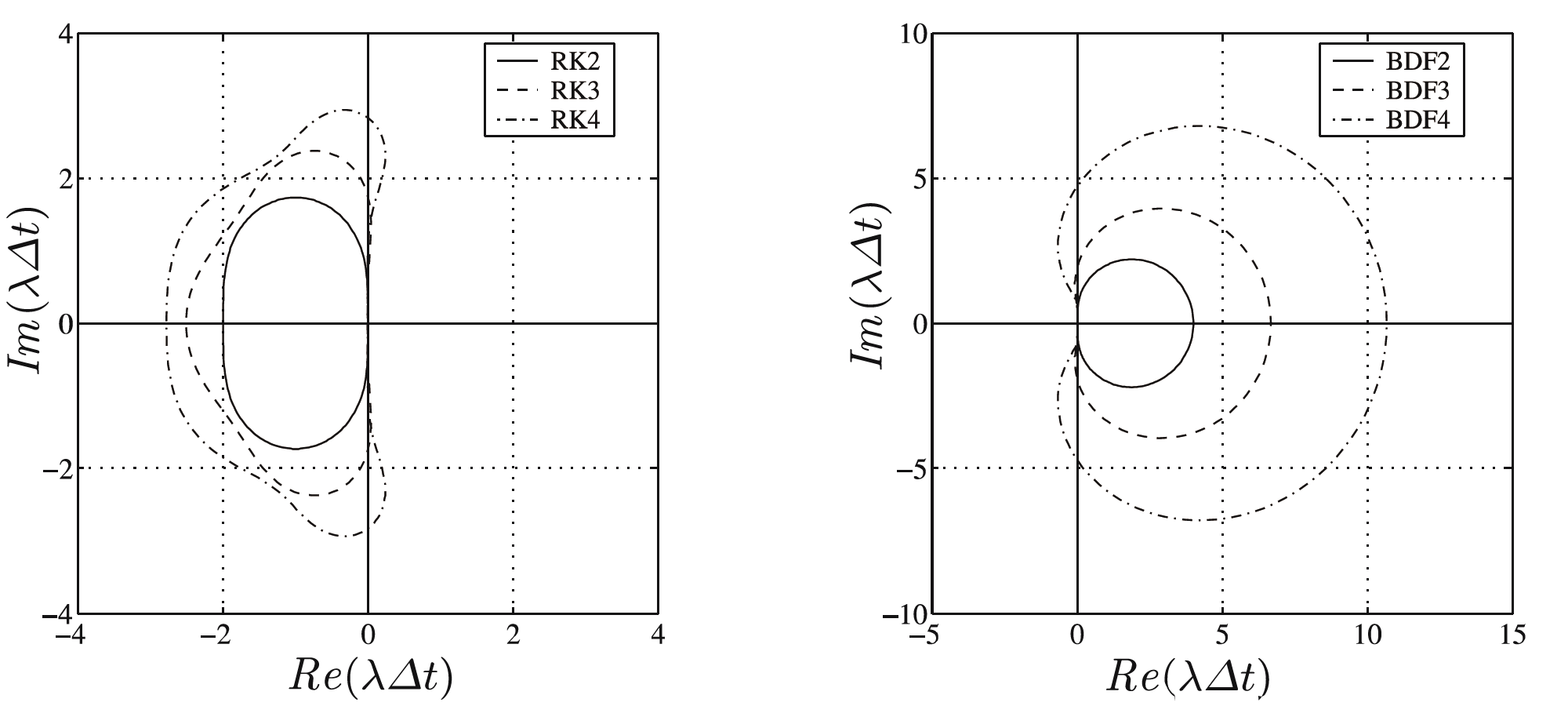}
  }
	\caption{Absolute stability regions of different time integration method. (\textit{a}) RK family schemes, (\textit{b}) BDF family schemes. The figure is from the literature of \citet{Canuto:2007}.}
	\label{fig:stabReg}
\end{figure}

Moreover, while the effectiveness of an explicit high-order scheme in reducing LES or DNS numerical dissipation is attractive, we should not ignore its related stability concerns from the commercial code and open source code. These low-dissipation schemes may still lead to divergence of results even if the common CFL conditions are met \citep{Canuto:2007} This instability is an essential property of these time-integral schemes and even shows up in solving ordinary differential equations (ODE). As shown in Fig.\ref{fig:stabReg}, when solving a particular problem, each time integral scheme has an absolute stability region in the complex plane. The solution is unconditionally stable only when the product of eigenvalue and step size falls within the stable region. Tab.\ref{tab:stabReg} shows the intersection of various method stable regions with negative real and positive imaginary axes. It can be seen that the three default schemes employed by the standard OpenFOAM release, the first-order backward Euler scheme (termed Euler), the second-order Crank--Nicolson scheme, and the BDF2 scheme (termed backward), are unconditionally stable. The Adams--Bashforth and Adams-Moutton families of linear multi-step method and the RK familiy of linear single-step method have their own stable intervals. Therefor, in order to ensure the stability of the calculation, the only way is to keep shrinking the step size. However, this may lead to an undesired and huge increase in the computational cost, reducing the efficiency of high-order method for precision improvement. This is why related researches should not stop at a single type/family or use only one particularly high-order scheme. For complex flow problems, it is almost impossible to calculate in advance the critical step size that is guaranteed to be stable, which makes it difficult to theoretically evaluate the performance of different schemes. A feasible solution is to use different low dissipation schemes for specific flow problems and compare their performance directly. Therefore, a low-dissipative universal solution framework which can dynamically select different schemes is necessary for the further study of scheme stability and real computational efficiency.

\begin{table}
  \caption{The intersection of the scheme stability region with the negative real axis and positive imaginary axis. The data comes from the literature of \citet{Canuto:2007}.}
  \centering
  \begin{tabular}{crl}
    \toprule
    Method            & $A \cap \mathbb{R}_{-}$ & $A \cap \mathbb{R}_{+}$ \\
    \midrule
    Leap frog         & $\{0\}         $        & $[0, 1]      $          \\
    Forward Euler     & $[-2, 0]       $        & $\{0\}       $          \\
    Crank--Nicolson   & $(-\infty, 0]  $        & $[0, +\infty)$          \\
    Backward Euler    & $(-\infty, 0]  $        & $[0, +\infty)$          \\
    \midrule
    AB2               & $(-1, 0]       $        & $\{0\}       $          \\
    AB3               & $[-6/11, 0]    $        & $[0, 0.723]  $          \\
    AB4               & $[3/10, 0]     $        & $[0, 0.43]   $          \\
    AM3               & $[-6, 0]       $        & $\{0\}       $          \\
    AM4               & $[-3, 0]       $        & $\{0\}       $          \\
    \midrule
    BDF2              & $(-\infty, 0]  $        & $[0, +\infty)$          \\
    BDF3              & $(-\infty, 0]  $        & $[0, 1.94)   $          \\
    BDF4              & $(-\infty, 0]  $        & $[0, 4.71)   $          \\
    \midrule
    RK2               & $[-2, 0]       $        & $\{0\}       $          \\
    RK3               & $[-2.51, 0]    $        & $[0, 1.73]   $          \\
    RK4               & $[-2.79, 0]    $        & $[0, 2.83]   $          \\
    \bottomrule
  \end{tabular}
  \label{tab:stabReg}
\end{table}

In this study, a low-dissipative universal solution framework using projection method was implemented based on the OpenFOAM libraries. The unified solver obtained by this framework can dynamically select a variety of low-dissipative high-order schemes in the runtime based on the case settings, including linear multi-step method and linear single-step method. The significance of this solution framework is as follows: a) a unified low-dissipative projection solver offers more choices for researchers who are interested in simulation of practical engineering problems and help to compare the real performance of different high-order method in specific flow problems; b) a general framework which is similar to the native solver is convenient for researchers who are interested in the numerical method to implement a variety of high-order and low-dissipation schemes and carry out their own scheme design for improvement; c) universal framework support for OpenFOAM native sources (termed fvOptions), turbulence models and other classes makes it applicable to a wider range of problems, which contribute to the evaluation of various existing algorithms in specific flow type.

The objectives of this paper are to (1) provide a low-dissipative projection framework that can dynamically decide the scheme in run time and is compatible with OpenFOAM's native architecture and demonstrate its usage, for which two linear multi-step family, Adams--Bashforth and Adams--Bashforth--Moutton schemes, in addition to Runge--Kutta family of single-step method, are implemented to exhibit how to introduce new time discretization methods in the general framework, (2) comprehensively evaluate the advantages and disadvantages of low-dissipative solvers compared with traditional implicit solvers from the perspective of computational accuracy and computational speed through a series of benchmark cases, and (3) suggest how to better select low-dissipation schemes for scale-resolving turbulence simulation in industrial applications by studying the differences in accuracy and calculation speed of different method empolyed.

The structure of this paper is organized as follows: Section \ref{sec:theory} of this paper focuses on the theoretical approach behind the low-dissipative framework, and Section \ref{sec:implement} focuses on the implementation of the framework itself and the integration of multiple time schemes. Section \ref{sec:setup} briefly introduces the settings and purposes of different test cases, and Section \ref{sec:results} shows the numerical results then analyzes and discusses several key issues.

\section{Numerical theory}
\label{sec:theory}

\subsection{Governing equation}

In this paper, incompressible flow solver is taken as an example to implement the low-dissipative framework. The fluid mechanics problem dealt with is described by the following conservative equation:

\begin{align}
  \label{eq:NS:continuity}
  u_{j,j} &= 0,  \\
  \label{eq:NS:momentum}
  \partial_t u_i + (u_i u_j)_{,j} &= -p_{,i} + \sigma_{ij,j} + f_i,
\end{align}

where $p$ is the pressure over density (the dimension is $\mathrm{m^2/s^2}$) and $\sigma_{ij}$ is the viscous stress tensor which differs in form according to constitutive relations. $f_i$ represents a generalized source term that cannot be included in the other four terms and, depending on the specific problem, can be expressed as a driving (volume) force, a buoyancy force, or a inertial force in a rotating coordinate system.

\subsection{Projection method}
\label{ssec:projection}

For the high-order time integration scheme with low dissipation, there is no restriction on whether the flow is compressible. In fact, it is easier to implement high-order time discretization for many density-based Riemann solvers with explicit advance. However, due to the difference in types of partial difference equation (PDE) between low-and high-speed flows (elliptic or hyperbolic problems), low-dissipative high-order schemes need to be used with different pressure/density-velocity coupling algorithms. This research is mainly based on incompressible flow solvers to implement various low dissipation schemes and complete corresponding verification. The focus of the current general framework is on the pressure-velocity coupling algorithm. SIMPLE family and PISO family are common coupling algorithms that use implicit time discretization to solve equations. Among them, PISO family algorithms usually have better performance under small time steps \citep{Issa:1986:1,Issa:1986:2}. Therefore, this algorithm is used to calculate the implicit time schemes for comparison in this paper. In the low-dissipative unified solver, the classical projection method of pressure-velocity coupling combined with explicit step advancing is employed in this study. There have been a lot of studies discussing the implementation of projection algorithm in OpenFOAM, and the specific details can be referred to the literature \citep{Vuorinen:2014}. The following is a brief explanation of the general procedure.

According to the Chorin's original version \citep{Chorin:1967,Chorin:1968}, the central idea of projection method is based on Helmholtz decomposition, that is, any vector field can be decomposed into irrotational and solenoidal components. Taking first-order accurate forward Euler method as an example, the procedure of the projection method is as follows. First, the intermediate velocity field is obtained explicitly according to the following equation:

\begin{equation}
\label{eq:predict}
  \frac{u_i^{*} - u_i^{n}}{\Delta t} = F_i \left( x_{(i)}, t, u_{(i)} \right),
\end{equation}

where

\begin{equation}
  F_i \left( x_{(i)}, t, u_{(i)} \right) = -(u_i u_j)_{,j} + \sigma_{ij,j} + f_i.
\end{equation}

The intermediate velocity field does not meet the divergence-free requirement, so it needs to be projected to the solenoidal component. By solving the pressure Poisson equation,

\begin{equation}
\label{eq:poisson}
  p_{,jj}^{n+1} = \frac{u_{j,j}^{*}}{\Delta t},
\end{equation}

the irrotational component $p_{,i}^{n+1}$ can be obtained. Finally, the final velocity field satisfying the divergence-free condition is obtained by the following projection (correction):

\begin{equation}
  u_i^{n+1} = u_i^{*} - p_{,i}^{n+1} \Delta t.
\end{equation}

Similar to the SIMPLE or PISO algorithm, the $u_i^{n+1}$ and $p^{n+1}$ obtained by one time projection does not strictly satisfy Eq.\eqref{eq:NS:momentum}, but when the time step is small enough or the discretization scheme's order of accuracy is high enough, the corresponding solution is considered to satisfy the momentum equation. Therefore, the projection method often needs to be used in combination with small time steps and explicit high-order methods, and it is naturally suitable for solution framework adopting low-dissipative time schemes.

\subsection{Linear single-step method}
\label{ssec:single}

Because of its advantages of achieving high order accuracy without additional storage of preceding time steps, the linear single-step method is favored by various CFD codes in the past when computer memory was limited. Among them, methods of Runge--Kutta family become the first choice for implementation of low-dissipative time integration method because of their simple form \citep{Komen:2021,Vuorinen:2014}. In order to facilitate the comparison of numerical accuracy and computational speed, two versions of the classical Runge--Kutta method, the Kutta version three-stage, third-order method (RK3) and the classical four-stage, forth-order method (RK4) are also implemented as examples in the low-dissipative framework. Readers can easily follow these two examples for more kinds of linear single-step implementation, such as the TVD Runge--Kutta method \citep{Shu:1988:2}.

The general form of the Runge--Kutta family is

\begin{equation}
  y_{n+1} = y_n + h \sum_{i=1}^{s} b_i k_i,
\end{equation}

where

\begin{equation}
  k_i = f \left( t_n + c_i h, y_n + h \sum_{j=1}^{s} a_{ij} k_j \right).
\end{equation}

The coefficient table of the three-stage and third-order Kutta method is as follows:

\begin{center}
  \begin{tabular}{c|ccc}
    0   & 0   & 0   & 0   \\
    1/2 & 1/2 & 0   & 0   \\
    1   & -1  & 2   & 0   \\ \hline
        & 1/6 & 2/3 & 1/6
  \end{tabular}
\end{center}

The coefficient table of the classical four-stage and forth-order Runge--Kutta method is as follows:

\begin{center}
  \begin{tabular}{c|cccc}
    0   & 0   & 0   & 0   & 0   \\
    1/2 & 1/2 & 0   & 0   & 0   \\
    1/2 & 0   & 1/2 & 0   & 0   \\
    1   & 0   & 0   & 1   & 0   \\ \hline
        & 1/6 & 1/3 & 1/3 & 1/6
  \end{tabular}
\end{center}

\subsection{Linear multi-step method}
\label{ssec:multi}

Unlike the single-step method, the linear multi-step method often require the solution of multiple previous time steps that have been calculated, which makes the implementation of the multi-step method often more complex than that of the single-step method. However, the accuracy of the multi-step method is usually higher, and because it does not need to perform multiple repeated calculations to obtain the required data as in the single-step method (the solution of previous time steps has already been acquired), it has an advantage in computational speed. Therefore, some in-house codes aimed at low dissipation and high-resolution turbulence simulation of complex flow problems prefer the multi-step method \citep{Chain:2015}.

The solution framework in this study contains two major linear multi-step method families, Adams--Bashforth series and Adams--Bashforth--Moutton series. The former is an explicit method, and the latter is a semi-implicit method of predictor-corrector scheme. As examples, the second-order Adams--Bashforth scheme (AB2) and the third-order Adams--Bashforth--Moutton method (ABM3) are implemented, and other researchers can easily imitate to achieve higher-order multi-step schemes as needed.

In general, linear multi-step methods can be expressed in the following form:

\begin{equation}
  \sum_{i=0}^{k} \alpha_i y_{n+i} = h \sum_{i=0}^{k} \beta_i f_{n+i},
\end{equation}

In particular, if the selection of $f_{n+i}$ is restricted, a more common form can be obtained:

\begin{equation}
  \begin{aligned}
    y_{n+k} &= y_{n-j} + h \sum_{i=0}^{q} \beta_{qi} f_{n-i},  \\
    \beta_{qi} &= \int_{-j}^k \prod_{l = 0, l \neq i}^{q} \frac{s+l}{-i+l} \mathrm{d}s.
  \end{aligned}
\end{equation}

\subsubsection{Adams--Bashforth method}

Let $k=1, j=0$ in (10), the explicit methods of Adams--Bashforth family can be obtained as

\begin{equation}
  y_{n+1} = y_n + h \sum_{i=0}^q \beta_{qi} f_{n-i}.
\end{equation}

The method obtained by the equations has a theoretical accuracy of $q+1$ order. For an Adams--Bashforth method of $q+1$ order, it is necessary to use the solution of the previous $q+1$ step. The first-order Adams--Bashforth scheme is the forward Euler scheme. The formula of second-order Adams--Bashforth scheme is

\begin{equation}
\label{eq:AB2}
  y_{n+1} = y_n + \frac{h}{2}(3f_n - f_{n-1}).
\end{equation}

\subsubsection{Adams--Moutton method}

Let $k=0, j=1$ in (10), the implicit methods of the Adams--Moutton family can be obtained as

\begin{equation}
  y_{n+1} = y_n + h \sum_{i=0}^q \beta_{qi} f_{n+1-i}.
\end{equation}

As can be seen from the equation, in addition to directly using the solution of the previous $q$ steps, it is necessary to solve the equation set implicitly to obtain the solution of the $n+1$ step. The method obtained above has a theoretical accuracy of $q+1$ order. The first-order Adams--Moutton method is the first-order backward Euler scheme, and the second-order Adams--Moutton method is the classic Crank--Nicolson scheme. Here is the specific form of the third-order Adams--Moutton method

\begin{equation}
\label{eq:AM3}
  y_{n+1} = y_n + \frac{h}{12}(5f_{n+1} + 8 f_n - f_{n-1}).
\end{equation}

Since the implicit scheme tends to introduce more dissipation, and the linear equation set solver required will increase the computation time of single step advance, the present universal framework does not directly implement Adams--Moutton method. Instead, the Adams--Bashforth--Moutton method, which incorporates a predictor-corrector procedure, is chosen.

\subsubsection{Adams--Bashforth--Moutton method}

The Adams--Moutton method needs the data of the unsolved time step, so the calculation process includes matrix solution. An alternative is to use the next time solution estimated by other explicit schemes instead for multi-step computation. This is the basic idea of predictor-corrector type schemes, that is, an explicit scheme and an implicit scheme are combined to obtain a semi-implicit method. By combining the Adams--Bashforth and Adams--Moutton family, a semi-implicit method of Adams--Bashforth--Moutton family can be obtained. Its common form is

\begin{equation}
  \begin{aligned}
    \widetilde{y}_{n+1} &= y_n + h \sum_{i=0}^q \beta_{qi} f_{n-i}, \\
    \widetilde{y}_{n+1} &= y_n + h \left( \beta_{q0} \widetilde{f}_{n+1} + \sum_{i=1}^q \beta_{qi} f_{n+1-i} \right).
  \end{aligned}
\end{equation}

The first-order Adams--Bashforth--Moutton method is the classical predictor-corrector Euler method. Here is the specific form of the adopted third-order Adams--Bashforth--Moutton scheme:

\begin{equation}
  \begin{aligned}
    \widetilde{y}_{n+1} &= y_n + \frac{h}{2} (3f_n - f_{n-1}), \\
    y_{n+1} &= y_n + \frac{h}{12} (5\widetilde{f}_{n+1} + 8f_n - f_{n-1}).
  \end{aligned}
\end{equation}

\section{Implementation details}
\label{sec:implement}

There is a lot of online tutorials or technique notes about the basic concepts related to OpenFOAM libraries, thus this part will not be discussed in this article. Readers can refer to the related study \citep{Vuorinen:2014} or the official manual \citep{FOAM} for primitive details. The major considerations of the low-dissipative projection framework implemented are as follows: 1) Employ a similar implementation style as the OpenFOAM native solver, so as to widely support the existing class of source terms, turbulence models or related controls. 2) Follow the OpenFOAM's design philosophy and build up scheme's run time selection. For specific cases, the adopted time scheme can be dynamically determined and solved according to a single file setup, without recompiling or designing a new solver. 3) The solution framework not only separates the general procedure from the special implementation of each scheme, but also separate the time discretization, pressure-velocity coupling algorithm, and linear equation matrix solution from the top level, so as to facilitate the subsequent introduction of other low-dissipative high-order methods or change of coupling algorithms. 4) The algorithms related to high-order time schemes are divided into several specific modules according to their different functions, and the newly added scheme only need to modify the corresponding modules according to the framework's basic ideas, avoiding redundant adjustments or designing solvers from the beginning for scheme-independent parts of the solving architecture.

The low-dissipative universal solution framework is implemented based on the above ideas, and a unified solver is finally compiled. When using the solver, users only need to change the keyword of a control file in the case directory, and can switch various high-order time scheme by selecting keyword similar to OpenFOAM's native scheme file. On the one hand, this avoids the extra work of having to write a new solver for adding a new scheme. On the other hand, it makes the comparison of results between different schemes eliminate the interference of other irrelevant factors of the solver (such as data accessing or field operations). Moreover, the framework includes both linear single-step schemes and linear multi-step schemes, which is conducive to evaluating the numerical dissipation and computational efficiency of different types of time discretization schemes in practical fluid mechanics problems, and helping to find out the most suitable low-dissipative scheme for specific flow type. Due to the separation in functions, other researchers can easily extend and embed other time integration method based on this common architecture or develop new advance schemes to further evaluate its performance and explore possible improvement of low-dissipative time schemes. The unified solver obtained by the framework maintains the support of various source terms, turbulence model and constitutive relation model of the standard OpenFOAM release. This means that other researchers using OpenFOAM for CFD calculations can switch solvers directly without having to make extensive compatibility changes to the solvers (such as for buoyancy or inertia forces) to suit their cases.

\subsection{Basic structure of the general framework}

\begin{figure}
	\centering
  \fbox{
    \includegraphics[width=10cm]{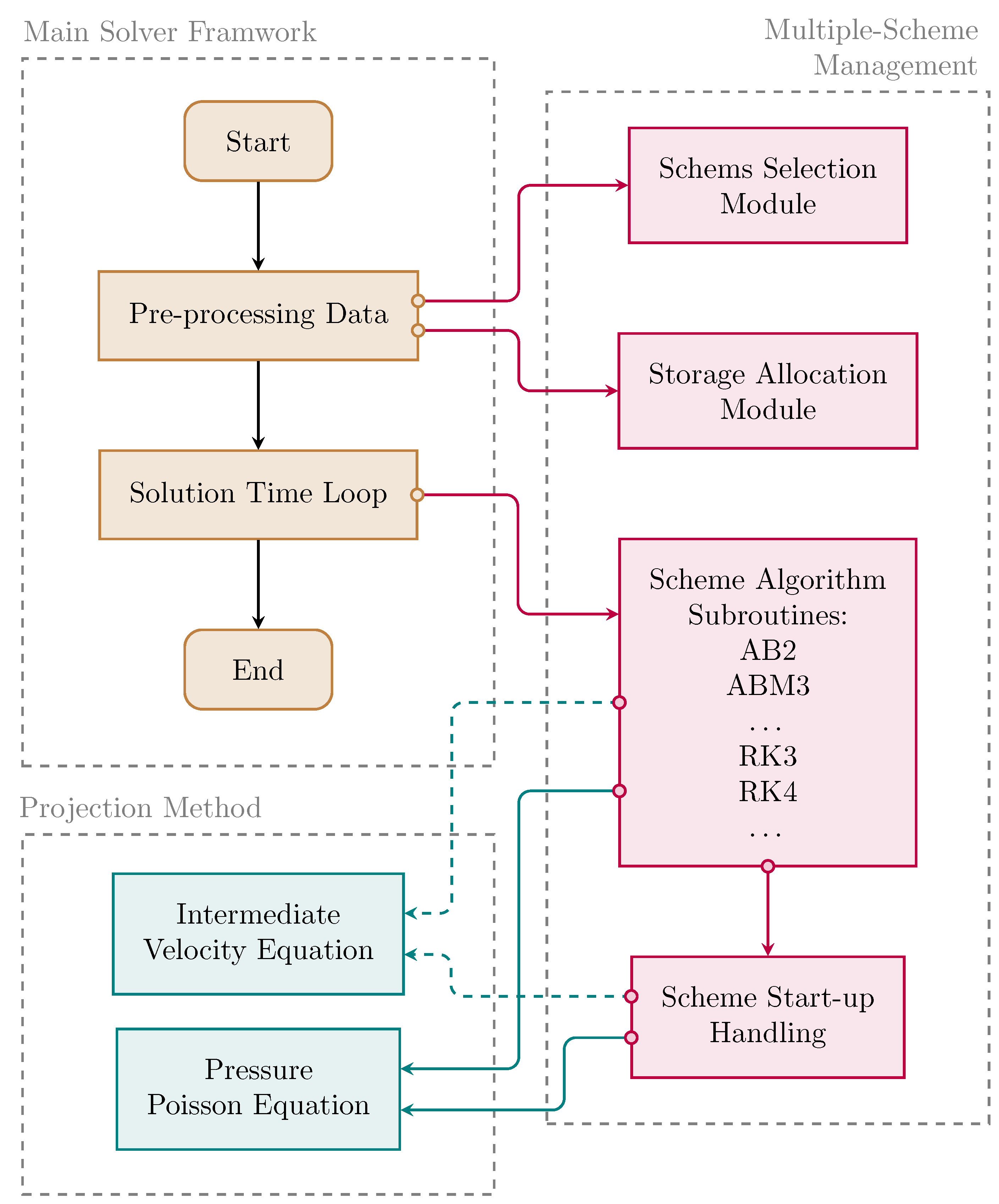}
  }
	\caption{Structure of the universal low-dissipative projection framework.}
	\label{fig:structure}
\end{figure}

The low-dissipative solution framework basic structure of the projection method implemented is shown in Fig.\ref{fig:structure}. Through this framework, the compiled unified solver fully retains the common form and style used in the high-level programming of standard OpenFOAM solvers, thus retaining support for native control methods, turbulent models, and various source terms. The brown color part in Fig.\ref{fig:structure} is the original architecture of the native solver. The extension of the low-dissipative framework is mainly achieved through the introduction of multiple-scheme processing modules at specific periods within the original architecture, which is identified in the red box in the figure. Among these multiple-scheme modules, the equation solving module which is the most important part for low-dissipative framework to realizes high-order time discretization, needs to use the projection algorithm to solve the pressure-velocity coupling. However, this part of the algorithm is independent of the specific higher-order scheme (for example, this algorithm is not required in the compressible solver), so in the design of the general framework, it is isolated and processed separately waiting the algorithm code pieces of high-order scheme to call directly if needed. The separation and invocation of the corresponding modules can be achieved through interfaces and dynamic libraries, or accomplished alternatively using multiple code pieces in ``*.H" for simplicity which is preferred by standard OpenFOAM solvers. The latter one is employed in this study, so as to ensure that a relatively independent unified solver tool can be compiled and run directly.

In general, the framework consists of four parts: the Scheme Algorithm Subroutines corresponding to the method explained in \S\S\ref{ssec:single} and \ref{ssec:multi}, the Schemes Selection Module for choosing scheme dynamically, and the Storage Allocation Module for allocating additional memory related to specific high-order scheme, and the Start-up Handling section used to start up the multi-step calculation.

\subsection{Constitution and invocation of projection method}

Generally speaking, when the high-order scheme framework is combined with the projected pressure-velocity coupling algorithm for calculation, the steps in \S\ref{ssec:projection} may be repeated more than once. The specific repeat number and step details depend on the chosen time integration method and are not exactly the same. But the solution procedure of the pressure Poisson Eq.\eqref{eq:poisson} is consistent regardless of the single-step method or the multi-step method. For most single-step methods, the form of the velocity Eq.\eqref{eq:predict} is also consistent, so the two parts can be modularized to separate the function of the coupling algorithm.

For intermediate velocity calculation steps, you can use the following code block (UEqn.H):

\begin{lstlisting}[language=C++]
  MRF.correctBoundaryVelocity(U);

  fvVectorMatrix UEqn
  (
      fvm::ddt(U)
    + fvc::div(phi, U)
    + MRF.DDt(U)
  ==
      fvc::laplacian(turbulence->nuEff(), U)
    + fvc::div((turbulence->nuEff())*dev2(T(fvc::grad(U))))
    + fvOptions(U)
  );

  fvOptions.constrain(UEqn);

  solve(UEqn);
\end{lstlisting}

The RK family method can be directly turned to this code block to calculate the intermediate velocity field. There are some point should be noticed in the above code. First, turbulence model is introduced in the matrix assembling process. Second, the equations are established taking into account the possible usage of typical OpenFOAM source terms, such as MRF to deal with rotating coordinate systems and fvOptions to represent generalized source terms. Third, due to the requirements of the first two points, the current solution framework uses fvMatrix objects to establish equations, ensuring the correct use of underlying algorithms, such as the source term constrain function or solve function for linear equation system.

The following code pieces (pEqn.H) is used to solve the Poisson equation and complete the velocity projection:

\begin{lstlisting}
  fvScalarMatrix pEqn
  (
      fvm::laplacian(p)
  ==
      fvc::div(U)/runTime.deltaT()
  );

  pEqn.setReference(pRefCell, pRefValue);

  pEqn.solve();


  // Compute U^{n+1}

  U -= fvc::grad(p)*runTime.deltaT();
  U.correctBoundaryConditions();
  phi = fvc::flux(U);
\end{lstlisting}

Similarly, with the fvMatrix class to assemble matrix, the use of the low-level iterative solver algorithm can be inherited. In summary, UEqn.H and pEqn.H constitute the major modules of the projection method, and different low-dissipative high-order methods can invoke this algorithm modules multiple times. If the two modules are called only once continuously, the procedure completes a first-order forward Euler scheme. Usually, these two modules are sufficient for single-step methods; However, for the multi-step schemes, the pEqn part can be directly used, but the intermediate velocity calculation section often needs to be re-implemented accordingly.

\subsection{Dynamic selection of low-dissipative scheme in run-time computation}

As shown in the Fig.\ref{fig:structure}, the implementation of scheme selection module is contained in the pre-processing part of the solution framework. This module uses C++ enumerator type and OpenFOAM extended Enum class object to build a dynamic recognition table for optional schemas. The identifier name for solver code identification is defined in the instance timeSchems as follows:

\begin{lstlisting}
  enum timeSchemes
  {
      AB2,
      ABM3,
      RK3,
      RK4,
  };
\end{lstlisting}

In order to make each time scheme in the framework as OpenFOAM native time scheme which can be switched through the specific settings file in the case directory, an instance object timeSchemeName of the Enum template class is also defined, providing the name (string) that the user needs to enter when selecting the scheme:

\begin{lstlisting}
  const Enum<timeSchemes> timeSchemeName
  ({
      { timeSchemes::AB2, "AdamsBashforth2" },
      { timeSchemes::ABM3, "AdamsBashforthMoutton3" },
      { timeSchemes::RK3, "Kutta3" },
      { timeSchemes::RK4, "RungeKutta4" },
  });
\end{lstlisting}

If other researchers need to introduce new high-order schemes into the framework, the corresponding names should be added in timeSchemes and timeSchemeName.

After introducing the scheme table, the solver also needs to be configured correctly to read the settings file:

\begin{lstlisting}
  IOdictionary dict
  (
      IOobject
      (
          "projectionControl",
          runTime.system(),
          runTime,
          IOobject::MUST_READ
      )
  );

  timeSchemes scheme = timeSchemeName.get("projectionScheme", dict);
\end{lstlisting}

According to the above settings, the solver automatically looks up the system/projectionControl file during the case run time and dynamically selects the specified scheme in the calculation according to the keyword. For example, if the projectionControl file is set to

\begin{lstlisting}
  projectionScheme        AdamsBashforthMoutton3;
\end{lstlisting}

then the unified solver will set scheme as the corresponding keyword ABM3, and select the third-order Adams--Bashforth--Moutton method for subsequent calculation and solution.

\subsection{Solving algorithm inside the time loop}

The framework enables the solution of partial differential equations \eqref{eq:NS:continuity} and \eqref{eq:NS:momentum} using a variety of high-order time integration schemes by adding the code pieces of subroutine within the time loop:

\begin{lstlisting}
  switch (scheme)
  {
      case AB2:
      {
          #include "AB2.H"
          break;
      }
      case ABM3:
      {
          #include "ABM3.H"
          break;
      }
      case RK3:
      {
          #include "RK3.H"
          break;
      }
      case RK4:
      {
          #include "RK4.H"
          break;
      }
      default:
      {
          FatalErrorIn(args.executable())
              << "Scheme" << timeSchemeName[scheme] << " is not implemented yet"
              << exit(FatalError);
      }
  }
\end{lstlisting}

As can be seen from the above code, to add a new method only need to add a new entry and the corresponding implementation *.H file. The detailed implementation for various schemes is put inside separate files to ensure functionality independence and simplicity. The second-order Adams--Bashforth method and the third-order Adams--Bashforth--Moutton method is used as examples to show how to incorporate new multi-step schemes into the general solution architecture.

\subsubsection{Implementation of multi-step scheme AB2}
\label{ssec:realizeAB2}

Take the second-order method as an example to introduce the Adams--Bashforth family. First, the velocity equation is constructed according to Eq.\eqref{eq:AB2}. Due to the vast changes in the form of the equation, the general velocity equation module of the projection method cannot be used. Instead, corresponding form of velocity equation is constructed in AB2.H and the matrix is assembled accordingly:

\begin{lstlisting}
  MRF.correctBoundaryVelocity(U);

  fvVectorMatrix UEqn
  (
      fvm::ddt(U)
    + 1.5*
      (
          fvc::div(phi, U)
        + MRF.DDt(U)
        - fvc::laplacian(turbulence->nuEff(), U)
        - fvc::div((turbulence->nuEff())*dev2(T(fvc::grad(U))))
      )
    - 0.5*
      (
          fvc::div(phi, Un)
        + MRF.DDt(Un)
        - fvc::laplacian(turbulence->nuEff(), Un)
        - fvc::div((turbulence->nuEff())*dev2(T(fvc::grad(Un))))
      )
  ==
      fvOptions(U)
  );

  // Update Un for next time step
  Un = U.oldTime();

  fvOptions.constrain(UEqn);

  solve(UEqn);
\end{lstlisting}

The above section should be noted that since AB2 needs to use the information of the previous time step, it needs to use an array of additional variable field data Un and update this array at each time step accordingly. The additional storage/memory requirements required by the multi-step method also need to be allocated depending on the specific scheme, and the implementation of this feature is discussed in \S\ref{ssec:memory}. The above code also takes into account support for source terms.

The pressure Poisson equation solution part of AB2 is completely consistent with the conventional projection method, and it only needs to include the corresponding fragments as follows:

\begin{lstlisting}
  #include "pEqn.H"
\end{lstlisting}

\subsubsection{Implementation of multi-step scheme ABM3}

This section takes the third-order method as an example to introduce the implementation of Adams--Bashforth--Moutton family. The implementation of ABM3 consists of two parts, namely AB2 and AM3 parts. The solution of all the pressure Poisson equations is exactly the same as the conventional projection method, and it only needs to call the corresponding pressure equation solving module.

Firstly, the calculation of AB2 scheme is carried out according to Eq.\eqref{eq:AB2}, and the corresponding steps are basically consistent with the steps in \S\ref{ssec:realizeAB2}. After the predicted velocity field is obtained, the AM3 scheme of Eq.\eqref{eq:AM3} is used for correction calculation. The corresponding implementation code is as follows:

\begin{lstlisting}
  const volVectorField& U_0 = U.oldTime();

  MRF.correctBoundaryVelocity(U);

  fvVectorMatrix UEqn
  (
      fvm::ddt(U)
    + 5.0/12.0
      *(
          fvc::div(phi, U)
        + MRF.DDt(U)
        - fvc::laplacian(turbulence->nuEff(), U)
        - fvc::div((turbulence->nuEff())*dev2(T(fvc::grad(U))))
      )
    + 8.0/12.0*
      (
          fvc::div(phi, U_0)
        + MRF.DDt(U_0)
        - fvc::laplacian(turbulence->nuEff(), U_0)
        - fvc::div((turbulence->nuEff())*dev2(T(fvc::grad(U_0))))
      )
    - 1.0/12.0*
      (
          fvc::div(phi, Un)
        + MRF.DDt(Un)
        - fvc::laplacian(turbulence->nuEff(), Un)
        - fvc::div((turbulence->nuEff())*dev2(T(fvc::grad(Un))))
      )
  ==
      fvOptions(U)
  );

  Un = U.oldTime();

  fvOptions.constrain(UEqn);

  solve(UEqn);
\end{lstlisting}

It should be noted in the above code that since ABM3 needs to use the data of the previous two time steps, in addition to using Un to store the data of the previous moment, it is also necessary to add the array U\_0 for the storage and usage of the current time step data.

\subsection{The startup of linear multi-step method}

Different from linear single-step method, linear multi-step method needs extra consideration of algorithm startup. This is because the multi-step method often needs solved data of multiple previous time step, but there is no sufficient step that meets the requirements of the high-order scheme at the zero moment. In the low-dissipative framework, scheme startup processing is included in different solving algorithm part of each scheme, and the corresponding startup methods are independently invoked in different schemes (as shown in the Fig.\ref{fig:structure}). This means that the solver can use different startup methods for different multi-step schemes as needed. The implementation of the startup module is completed by determining the serial number of the current time step:

\begin{lstlisting}
  if (runTime.timeIndex() == 1)
  {
      ...
  }
  else if (runTime.timeIndex() == 2)
  {
      ...
  }
  ...
  else
  {
      ...
  }
\end{lstlisting}

The number of time steps required for startup processing in the above code is determined by the algorithm related to specific method. Usually in previous researches, the single-step method of the same order is adopted to start the multi-step method to ensure that the order of accuracy is not reduced. One of the main objectives of this study is to compare the differences in numerical accuracy and computational speed of various low-dissipation schemes. Therefore, in order to ensure the independence of the algorithms of different schemes, the corresponding low-order method is used to start the multi-step method, for example, AB2 is started by AB1, ABM3 is started by ABM2.

\subsection{Dynamic memory allocation versus static storage reading and writing}
\label{ssec:memory}

Single-step and multi-step methods in high-order scheme often require additional arrays to store solved field data. For linear single-step methods, usually at least one array of the same size is required to store the accumulative results; For linear multi-step methods, several additional arrays are required to store the corresponding previous results. There are usually two ways to deal with the additional storage requirements of multi-step methods. The first approach is to limit the dynamic memory increase, but to store multiple field data statically on disk and read it when needed. This approach, which reduced memory burden but increased IO time to write/read from disk, was popular in the history when computer memory was much smaller than it is today. The second method is to put all the additional storage in dynamic memory, and do not perform static storage IO operations to save the corresponding IO time. From experience, in many of the servers, workstations, or HPC clusters that the authors have used, the hardware capabilities of the current computers are more than capable of bearing the increased memory burden of common high-order schemes. In the costly computations that the authors actually perform, the peak occupancy of dynamic memory in each computation node is often unsaturated. On the contrary, a large number of I/O operations will cause a large burden on the node (some clusters will impose a limit on the peak value of the user's I/O, and jobs that exceed this limit for a long time will be forced to suspend). Therefore, the low-dissipation solution framework in this study uses dynamic memory to store the additional array of field data, which is reflected in the added storage allocation module in the Figure 2. This module is also called in the data preprocessing module, but only after the solver has completed dynamic scheme selection. This means that the module can dynamically allocate the size of the storage according to the needs of different schemes to avoid possible waste during high-load calculations.

This section uses AB2 as an example to briefly introduce the dynamic storage allocation of high-order schemes:

\begin{lstlisting}
  autoPtr<volVectorField> aUn;

  MRF.correctBoundaryVelocity(U);

  aUn =
      autoPtr<volVectorField>::New
      (
          IOobject
          (
              "Un",
              runTime.timeName(),
              mesh,
              IOobject::NO_READ,
              IOobject::NO_WRITE
          ),
          U
      );
\end{lstlisting}

The above code dynamically allocates memory through OpenFOAM's pointer class autoPtr, which can free memory at any time as needed to reduce peak usage. High-order methods may require more dynamic pointers to manage the corresponding solution's data separately.

\subsection{Source term compatibility of the framework}

Some previous OpenFOAM based implementations of explicit high-order time-discretization solvers tend to omit the processing of source terms. OpenFOAM uses a special mechanism that use fvOptions instances to abstract and unify source terms with different mechanisms (such as meanVelocityForce or buoyancyForce). In practice, the new unified solver should be able to use the same mechanism as the OpenFOAM native solver to support a variety of existing source term types, otherwise researchers with specific source term requirements will have to write a new source term class specifically for the new solver or repeatedly adjust the matrix assembly in the solver source code. Therefore, the support for fvOptions related mechanisms is considered in both single-step and multi-step methods in this framework. Note that in the multi-step method, fvOptions are provided on the right-hand side of the equation in a same form. This is because OpenFOAM's built-in source term implementation only supports this operation. If the researchers want to include the source term calculation in the multi-step method strictly, a new source term class is inevitable.

\section{Details and setup of test cases}
\label{sec:setup}

In this paper, four test cases are used to study the numerical accuracy and computational efficiency of the low dissipation framework. There are five schemes involved, namely backward (second-order backward Euler method combined with PISO coupling algorithm), AB2 (linear multi-step method), ABM3 (linear multi-step method), RK3 (linear single-step method), and RK4 (linear single-step method). Case 1 first verifies that the practical order of accuracy of the implemented scheme reaches the theoretical one. Case 2 verifies the effectiveness of the unified solver for solving classical flow problems using various low-dissipation schemes, where the total computation time required for each scheme is compared. Case 3 uses the implicit LES method (ILES) to verify that the four schemes under the projection framework have lower numerical dissipation compared with PISO-backward method, and the calculation speed of this ILES case is analyzed as well. Finally, Case 4 further investigates the instability of explicit high-order schemes through a quasi-DNS simulation of the channel flow problem which lack of enough dissipation to keep the calculation stable in large time step. The results of this case illustrate the potential advantages and disadvantages of employing low-dissipative time-discretization solvers in high-resolution scale-resolving turbulence simulations.

The test cases involving parallel computing in this study are all completed on high performance computation cluster. Each computation node of the cluster consists of two Intel® Xeon® Platinum 9242 processors, for a total of 96 cores. The maximum memory of each node is 384G, and the maximum memory usage was not approached in the entire computation process. Except for Case 4, the benchmark calculations were performed in the same node (including some one-processor cases) and did not involve cross-node data exchange.

Since the main calculation time of the low-dissipation projection framework comes from solving Poisson equations, the detail settings of the iterative methods to solve linear equations are guaranteed to be completely consistent when comparing the computation speed/time of each high-order scheme.

\subsection{Case 1: order of accuracy verification}

Case 1 is constructed to verify the real convergence order of each time scheme. The following ordinary differential equation (ODE) is solved :

\begin{equation}
\label{eq:ODE}
  \begin{aligned}
    \frac{\mathrm{d}y}{\mathrm{d}t} &= -a y + b cos(\omega t),  \\
    y(0) &= \frac{ab}{a^2 + \omega^2}.
  \end{aligned}
\end{equation}

The above ODE has analytic solutions:

\begin{equation}
  y = \frac{b}{a^2 + \omega^2} [\omega \sin(\omega t) + a \cos(\omega t)].
\end{equation}

Considering that we need to the order of schemes that embedded in the low dissipation framework, we should not rewrite a special ODE program to do the calculation. Instead, we have accomplished this ODE problem in the conventional CFD procedure with the following setup: a) use the unified solver on a one-cell grid; b) all boundary conditions are set to Neuman boundary to eliminate spatial transport terms; c) establish the terms in the right-hand side of Eq.\eqref{eq:ODE} by setting the source term of the momentum equation. With these settings and given the corresponding initial field, governing equation \eqref{eq:NS:momentum} of framework/solver is converted to Eq.\eqref{eq:ODE}. Fixing the total time, the error of the solution result is calculated by using the resolution of 32, 64, 128 and 256 timesteps, and the corresponding real convergence order of accuracy is obtained accordingly.

\subsection{Case 2: lid-driven cavity}

Case 2 is the classical problem of lid-driven cavity (LDC), and the Reynolds numbers $\mathrm{Re} = 1000$ and $\mathrm{Re} = 2300$ are adopted. Case 2 uses grids of $50 \times 50$, $100 \times 100$, $200 \times 200$, and $400 \times 400$ resolutions, corresponding to parallel computing setups of 1, 4, 16, and 64 cores, respectively. This setup ensures that each process handles the same number of cells and the calculation speed metric used in this case is the number of time steps that can be advanced per unit CPU time.

\subsection{Case 3: decay of homogeneous and isotropic turbulence}

Case 3 calculates the decay of homogenous and isotropic turbulence (DHIT) using ILES. Instead of verifying the accuracy of framework, this case focus on the dissipation of the implemented schemes. In order to demonstrate scheme's dissipation, we did not adopt optimal settings aimed at obtaining results close to experimental data (for accurate verification results, please refer to the literature of \citet{Guo:2023:JFM}), but employ special strategies to reduce the physical (model) dissipation and other numerical dissipation in addition to the scheme dissipation. ILES chose the zero-equation sub-grid model WALE model. When the cell size gradually increases, the sub-grid model will not provide sufficient model dissipation, and the total dissipation of the numerical method (including model dissipation and numerical dissipation of the grid) will be insufficient, which causes turbulent energy spectrum to rise abnormally in small scales and violate the Kolmogorov energy cascade theory. Usually, this indicates a bad realization of ILES computation if the goal is to obtain accurate results. However, this phenomenon benefits our verification of scheme's low-dissipation property. This is because that the numerical dissipation of the scheme in the practical calculation are outstanding in this test case and can be verified by comparing the magnitude of the energy spectrum deviation: if the dissipation of the scheme is lower, the overall dissipation of turbulence in the ILES calculation is more insufficient, and the turbulent kinetic energy spectrum will show a greater abnormal rise in small scales.

In Case 3, three grid resolutions, $64^3$, $128^3$ and $256^3$, were adopted, and all resolutions use 64 cores. The time step of $64^3$ and $128^3$ is the same, while the time step of $256^3$ is halved. The computational speed metric used in Case 3 is the number of time steps that can be advanced per process per unit CPU time on average.

\subsection{Case 4: channel flow}

Case 4 is the benchmark of periodic channel flows (PCF). As mentioned above, in addition to the well-known advantages, the stability problem is the main drawback that restricts application of explicit time integration schemes. The purpose of this case is not to obtain accurate calculation results as well (for accurate verification results, please refer to the reference of \citet{Guo:2023:JFM}), but to further study the difference in results' accuracy and consuming time of each scheme implemented by the universal framework in view of this instability issue. In this case, the computational grid is exactly the same as in the literature on DNS research \citep{Kim:1987}, and a LES computation without sub-grid model (quasi-DNS) is carried out. Because the dissipation of the method is reduced to a very low level, the case is very unstable when using low-dissipative high-order time discretization. Even when the CFL conditions are clearly satisfied, the explicit scheme need to greatly reduce the time step to increase the stability to ensure that the calculation does not diverge and the critical time step that guarantees convergence depends on the specific scheme. Based on this setup, Case 4 compares the statistical turbulence correlations obtained by different schemes and the computation times required for each calculation stage.

The standard baseline case adopts backward scheme to perform 20s transition calculation and 80s statistical calculation, with time step of 0.0025s and corresponding maximum CFL number approximately 0.12. Under the premise of ensuring the transition computation time and statistical computation time, the four low-dissipative schemes are calculated with different time steps due to the limitation of computational stability. The time step corresponding to the AB2 scheme is 5e-5s, and the time step of ABM3, RK3, and RK4 is 1.25e-4s. It has been tested that further increasing the corresponding step size in this case will result in a divergence of the calculation in the corresponding scheme. In addition, another case is set to consider the effect of extending the calculation time period. The time-extended case also adopts backward scheme and corresponding time step size, but increases the transition computation time to 100s and the statistical computation time to 200s respectively.
In Case 4, 128 cores were used to complete the calculation. Each example uses a total of 2 computation nodes, ensuring the 96+32 configuration. During the calculation, the remaining memory of the two nodes is sufficient indicating that the memory capacity limit is not approached.

\section{Results and discussion}
\label{sec:results}

\subsection{Case 1: order of accuracy verification}

Table \ref{tab:order} shows the calculated convergence order based on $L_2$ error. As the resolution (the number of time steps) increases, the AB2 and backward scheme results converge to second-order precision, with the error decreasing slightly faster in AB2. ABM3 and RK3 both show the error convergence of third order, and RK4 exhibits the fourth order of accuracy. When the time resolution increases from 128 to 256, the convergence rate of RK4 decreases slightly. This is because the results of RK4 are highly accurate, and the corresponding error at 128 resolution is already very small (less than the error at 256 resolution of other schemes). This value is very close to the output precision set by the calculation (6 significant digits), and further reducing the step size will not reduce the error beyond the output precision due to this limitation.

\begin{table}
  \caption{Order of accuracy for different implemented schemes.}
  \centering
  \begin{tabular}{cccc}
    \toprule
    Scheme   & Mesh & $L_2$ Error & Order       \\
    \midrule
    backward & 32   & 0.186472753 &             \\
             & 64   & 0.057394844 & 1.699971784 \\
             & 128  & 0.015282578 & 1.909033228 \\
             & 256  & 0.003892437 & 1.973142426 \\
    \midrule
    AB2      & 32   & 0.427281144 &             \\
             & 64   & 0.072246279 & 2.564190463 \\
             & 128  & 0.018466548 & 1.968009088 \\
             & 256  & 0.004652828 & 1.988734511 \\
    \midrule
    ABM3     & 32   & 0.218402013 &             \\
             & 64   & 0.02024748  & 3.431171919 \\
             & 128  & 0.002254222 & 3.167041065 \\
             & 256  & 0.000267948 & 3.072602544 \\
    \midrule
    RK3      & 32   & 0.049595293 &             \\
             & 64   & 0.004795928 & 3.370321312 \\
             & 128  & 0.000524581 & 3.192573392 \\
             & 256  & 6.14244E-05 & 3.094280949 \\
    \midrule
    RK4      & 32   & 0.01041488  &             \\
             & 64   & 0.00051152  & 4.347711254 \\
             & 128  & 2.83478E-05 & 4.173481903 \\
             & 256  & 2.72938E-06 & 3.376589555 \\
    \bottomrule
    \end{tabular}
  \label{tab:order}
\end{table}

\subsection{Case 2: lid-driven cavity}

\begin{figure}
	\centering
  \fbox{
    \includegraphics[width=8cm]{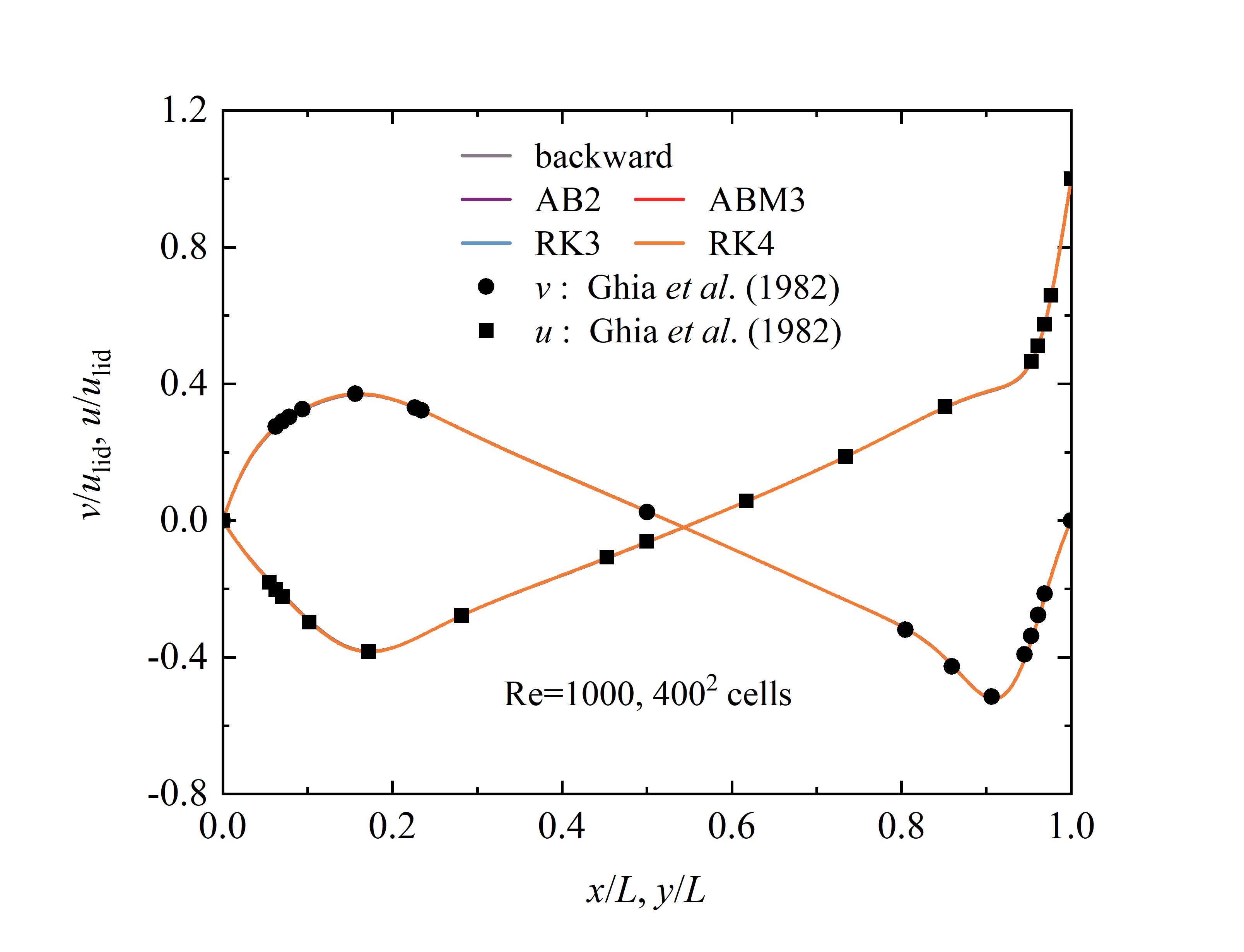}
    \includegraphics[width=8cm]{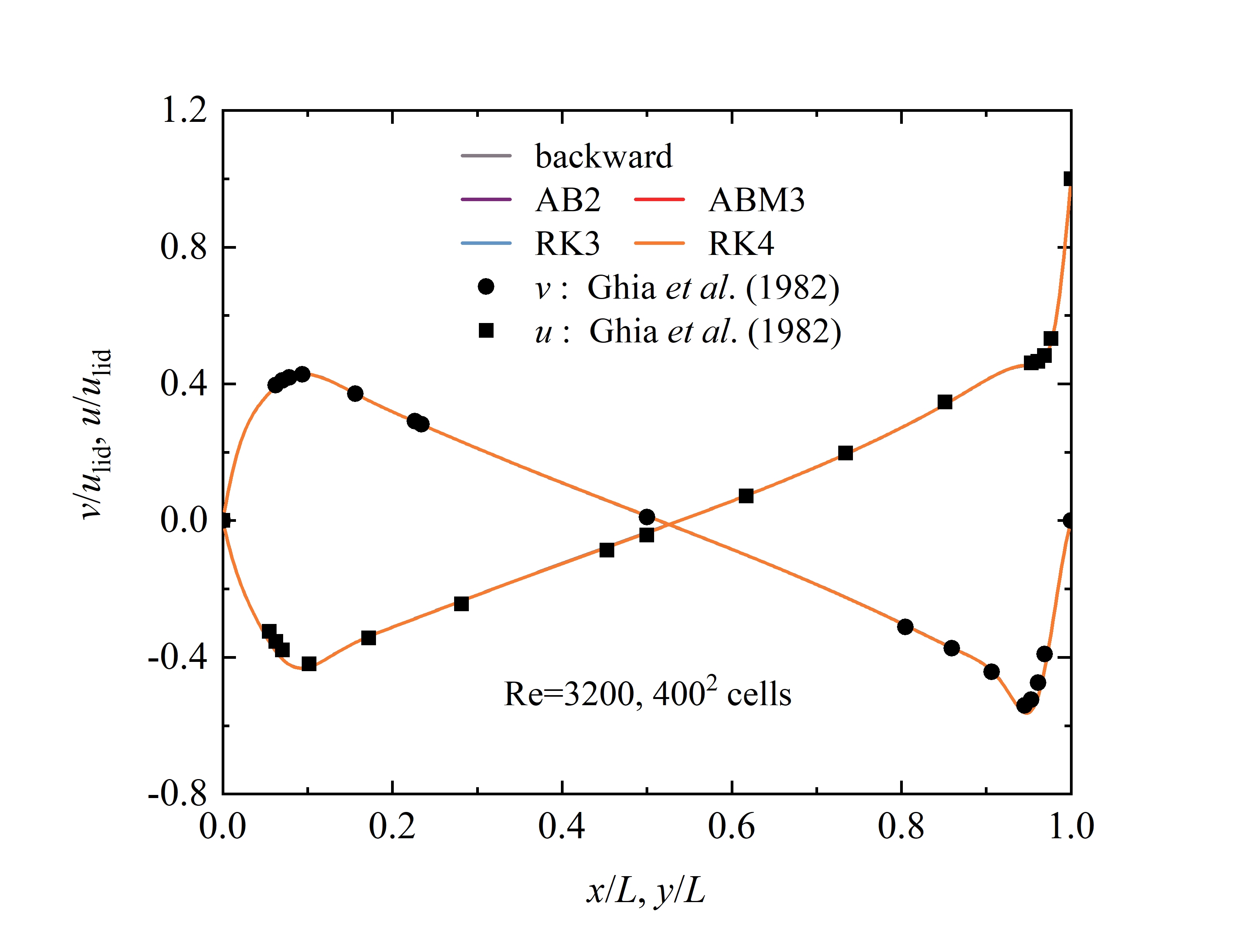}
  }
	\caption{Velocity profile of lid-driven problem at different Reynolds numbers. (\textit{a}) $\mathrm{Re}=1000$, (\textit{b}) $\mathrm{Re}=3200$. The reference data are from the literature of \citet{Ghia:1982}.}
	\label{fig:LDC:profile}
\end{figure}

Fig.\ref{fig:LDC:profile} shows the velocity profile distribution obtained by each time scheme. As can be seen from the figure, under the framework of low dissipation projection method, the calculated results of both linear single-step method and linear multi-step method are relatively consistent with the implicit backward-PISO result. All numerical results are consistent with the reference values. This illustrates the effectiveness of the low-dissipative framework and the two kinds of multi-step scheme implemented.

\begin{figure}
	\centering
  \fbox{
    \includegraphics[width=8cm]{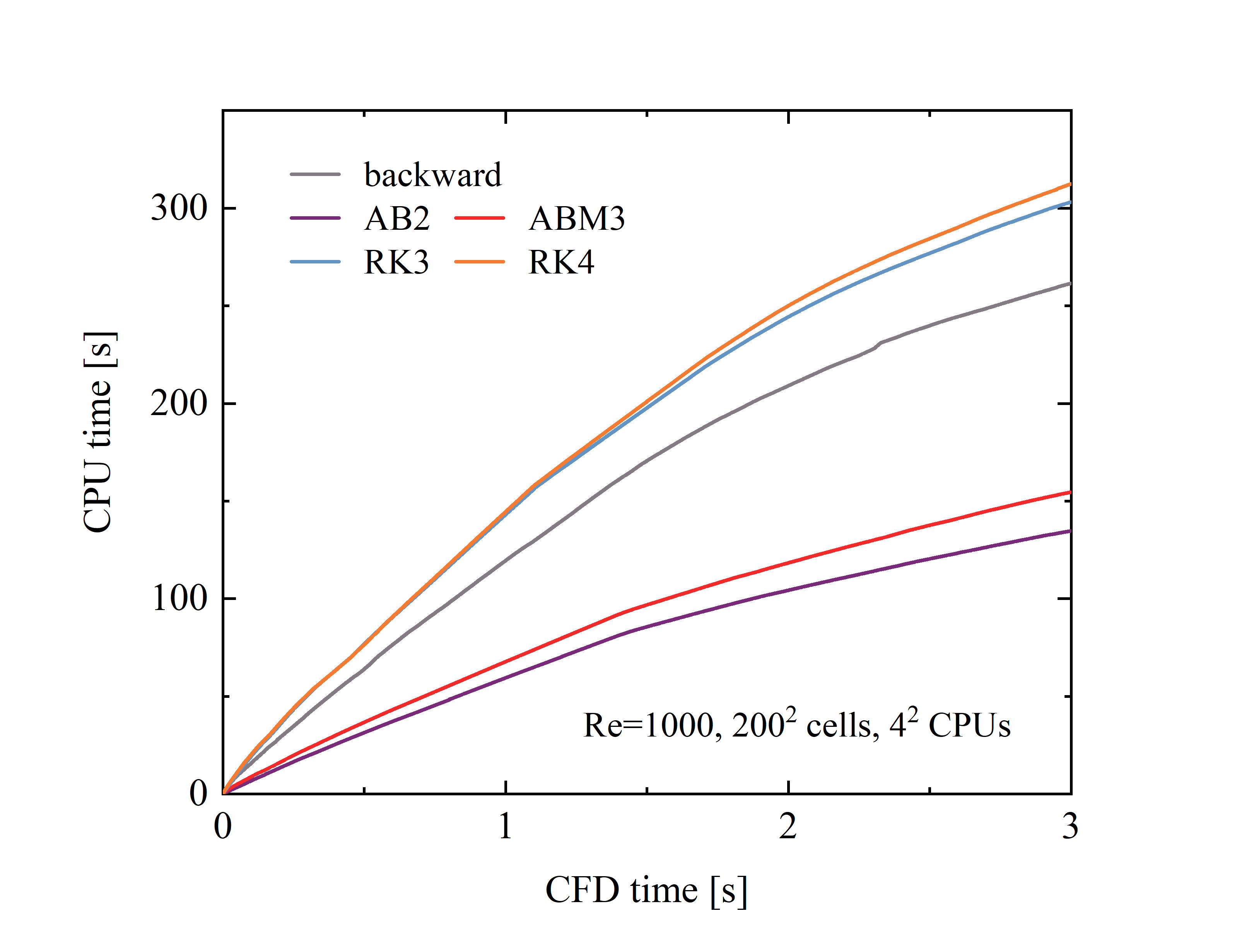}
    \includegraphics[width=8cm]{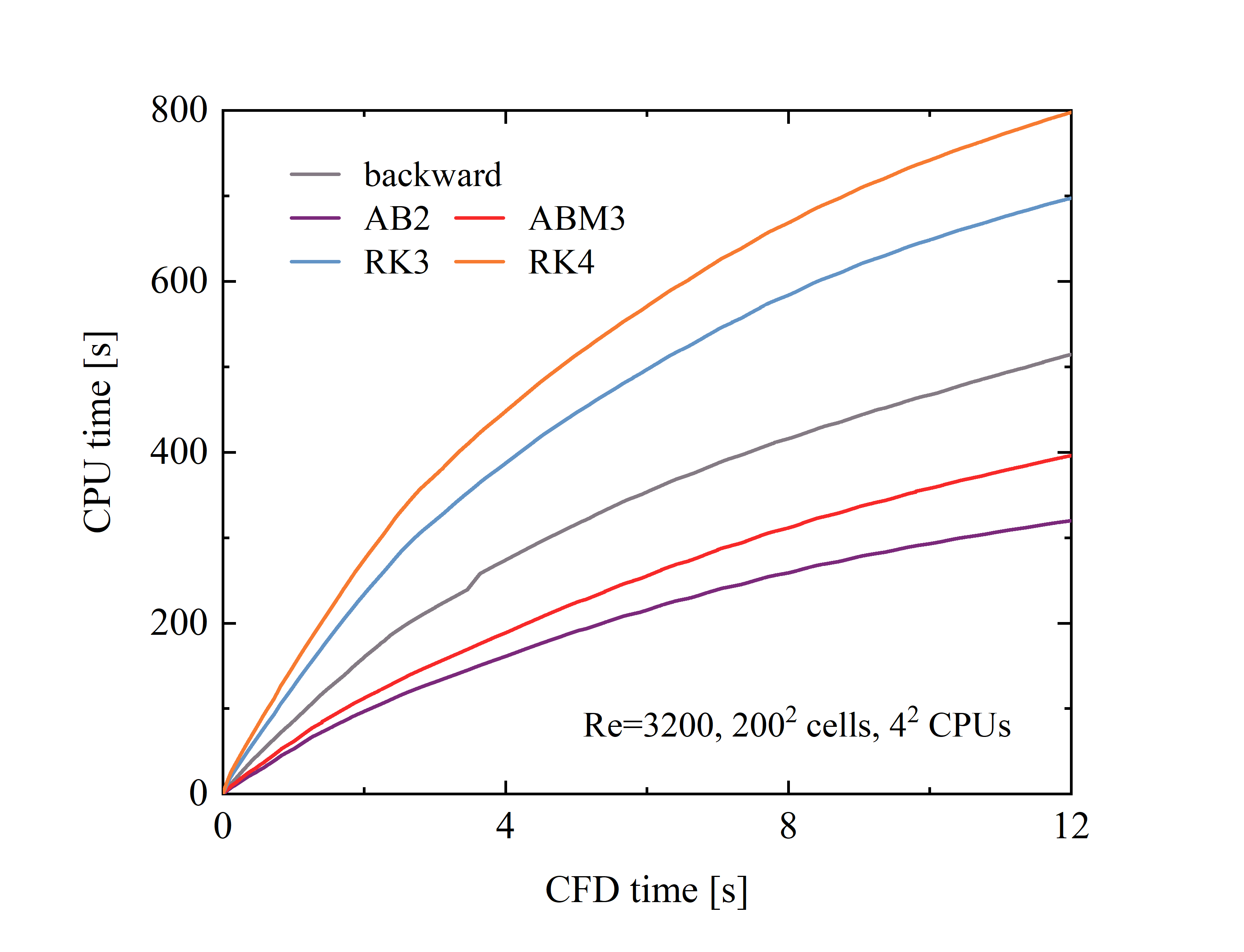}
  }
	\caption{Cumulative computation time (CPU time) of lid-driven problem at different Reynolds numbers. (\textit{a}) $\mathrm{Re}=1000$, (\textit{b}) $\mathrm{Re}=3200$.}
	\label{fig:LDC:speed}
\end{figure}

Although there is little difference between the accuracy of the various low-dissipation schemes and the backward scheme for the LDC problem shown in Fig.\ref{fig:LDC:profile}, there is a clear difference in calculation speed between these discretization methods. Fig.\ref{fig:LDC:speed} shows the increase in total CPU time over CFD time for the five schemes. At the beginning, the transient of the calculated physical quantities are significant, and it takes longer time to solve the passion equation. When the calculation begins to enter the quasi-steady state stage, the computation speed gradually increases. Under the condition of similar accuracy of calculation results, it can be seen from the figure that the calculation time of multi-step method is greatly reduced: second-order method AB2 and third-order method ABM3 are both faster than backward. This is partly due to the reduced cost of matrix solving in the explicit methods, and partly because linear multi-step methods increase the computation speed at the cost of storage (there is no need to perform multi-ple computations like single-step methods). Fig.\ref{fig:LDC:speed} also exhibits that RK3 does not save about 25\% computation time over RK4 as expected. This is because in the actual calculation, the matrix iteration often needs to reach the same residual level before terminated, which leads to the inconsistency of the iterative solving time of the Poisson equations in these two schemes.

In addition, the relation of the computation time between these five schemes may vary depending on the physical problem: for example, the time required for RK3 and RK4 is basically the same at $\mathrm{Re}=1000$, while a clearer difference in computation time can be observed at $\mathrm{Re}=3200$.

More detailed speed information for each grid number case are shown in Table \ref{tab:LDC:1000} and Table \ref{tab:LDC:3200} respectively. Although each core (process) computes the same number of grid cells, the table shows that as the number of cores increases, the computation speed decreases gradually. This shows that in the current problem, the increase in computing cost caused by inter-core data transmission is significant. Moreover, consistent with the results in Fig.\ref{fig:LDC:speed}, when $\mathrm{Re}=1000$, RK3 calculation speed is almost consistent with RK4 under different grid numbers. When $\mathrm{Re}=3200$, RK3 has an apparent speed increase compared with RK4. But with the increase of the number of grids, this speed increase is gradually covered by the time consumption of inter-core data transmission.

In general, the linear multi-step scheme in the low dissipation projection solution framework shows significant speed advantages in various case, especially the calculation speed of ABM3, which is nearly twice that of the same order scheme RK3. This shows that the multi-step scheme implemented in the framework can significantly improve the computation speed while the results differ little in accuracy.

\begin{table}
  \caption{Computation speed of LDC under different mesh resolution when $\mathrm{Re}=1000$.}
  \centering
  \begin{tabular}{cccccc}
    \toprule
    \multirow{2}{*}{Processes} & \multicolumn{5}{c}{Computation speed (steps per second)}            \\
    \cmidrule{2-6}
                               & backward    & AB2         & ABM3        & RK3         & RK4         \\
    \midrule
    1                          & 177.0956316 & 330.7607497 & 249.3765586 & 138.6962552 & 125.6281407 \\
    4                          & 93.70607528 & 175.9014952 & 137.4570447 & 69.60556845 & 65.02655251 \\
    16                         & 45.88032881 & 89.04719501 & 77.60460454 & 39.57783641 & 38.40245776 \\
    64                         & 25.4906959  & 53.37722127 & 45.55116915 & 23.76614117 & 22.99246997 \\
    \bottomrule
    \end{tabular}
  \label{tab:LDC:1000}
\end{table}

\begin{table}
  \caption{Computation speed of LDC under different mesh resolution when $\mathrm{Re}=3200$.}
  \centering
  \begin{tabular}{cccccc}
    \toprule
    \multirow{2}{*}{Processes} & \multicolumn{5}{c}{Computation speed (steps per second)}            \\
    \cmidrule{2-6}
                               & backward    & AB2         & ABM3        & RK3         & RK4         \\
    \midrule
    1                          & 281.0962755 & 468.75      & 319.8294243 & 187.4414246 & 160.7932467 \\
    4                          & 139.3809164 & 211.2490098 & 167.0843776 & 90.04614865 & 83.24373071 \\
    16                         & 93.29083418 & 150.0140638 & 121.160108  & 68.81917761 & 60.16394675 \\
    64                         & 47.81805231 & 81.30769882 & 70.39671482 & 39.60608448 & 38.37850804 \\
    \bottomrule
    \end{tabular}
  \label{tab:LDC:3200}
\end{table}

\subsection{Case 3: decay of homogeneous and isotropic turbulence}

\begin{figure}
	\centering
  \fbox{
    \parbox[b]{16cm}{
      \centering
      \includegraphics[width=7.9cm]{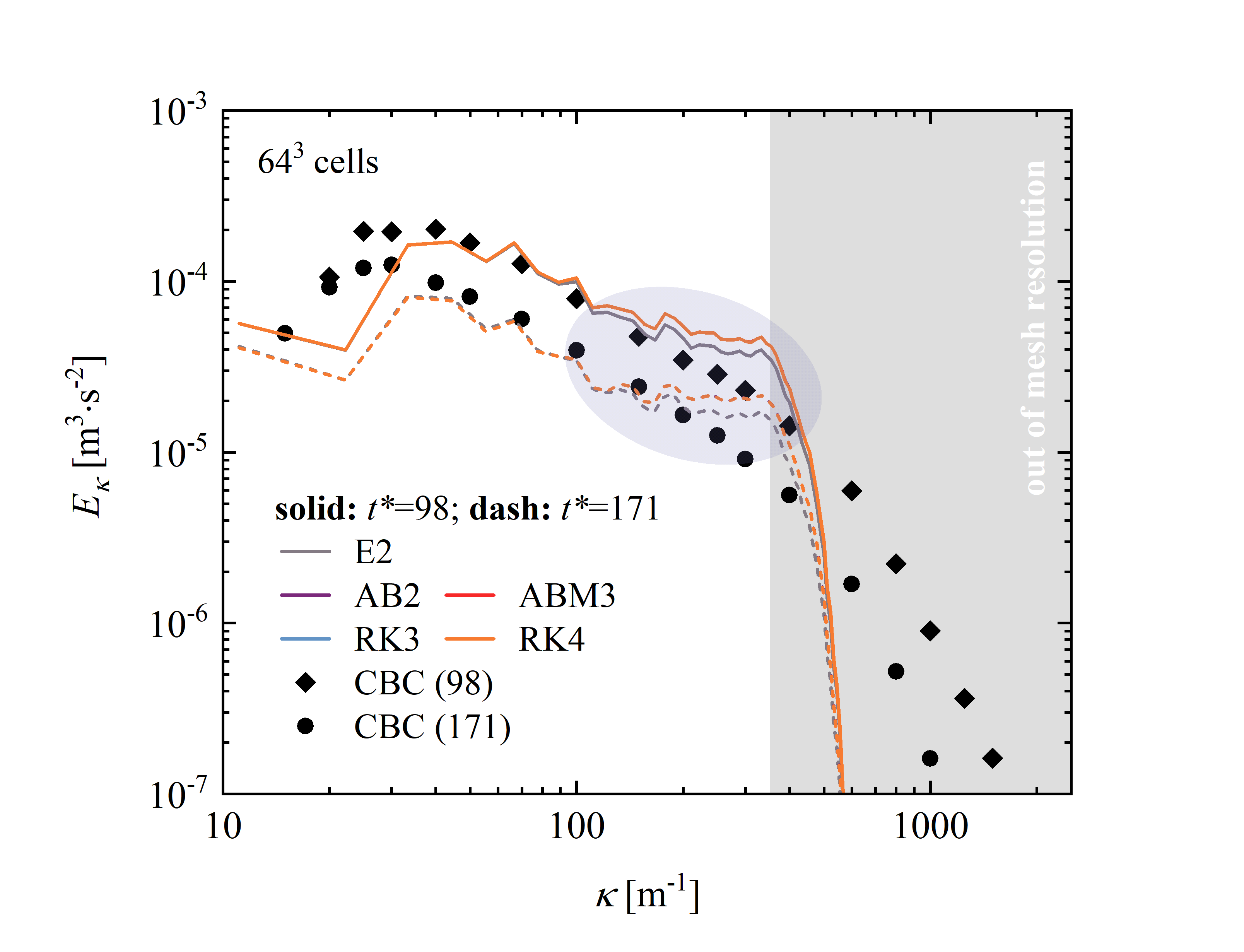}
      \includegraphics[width=7.9cm]{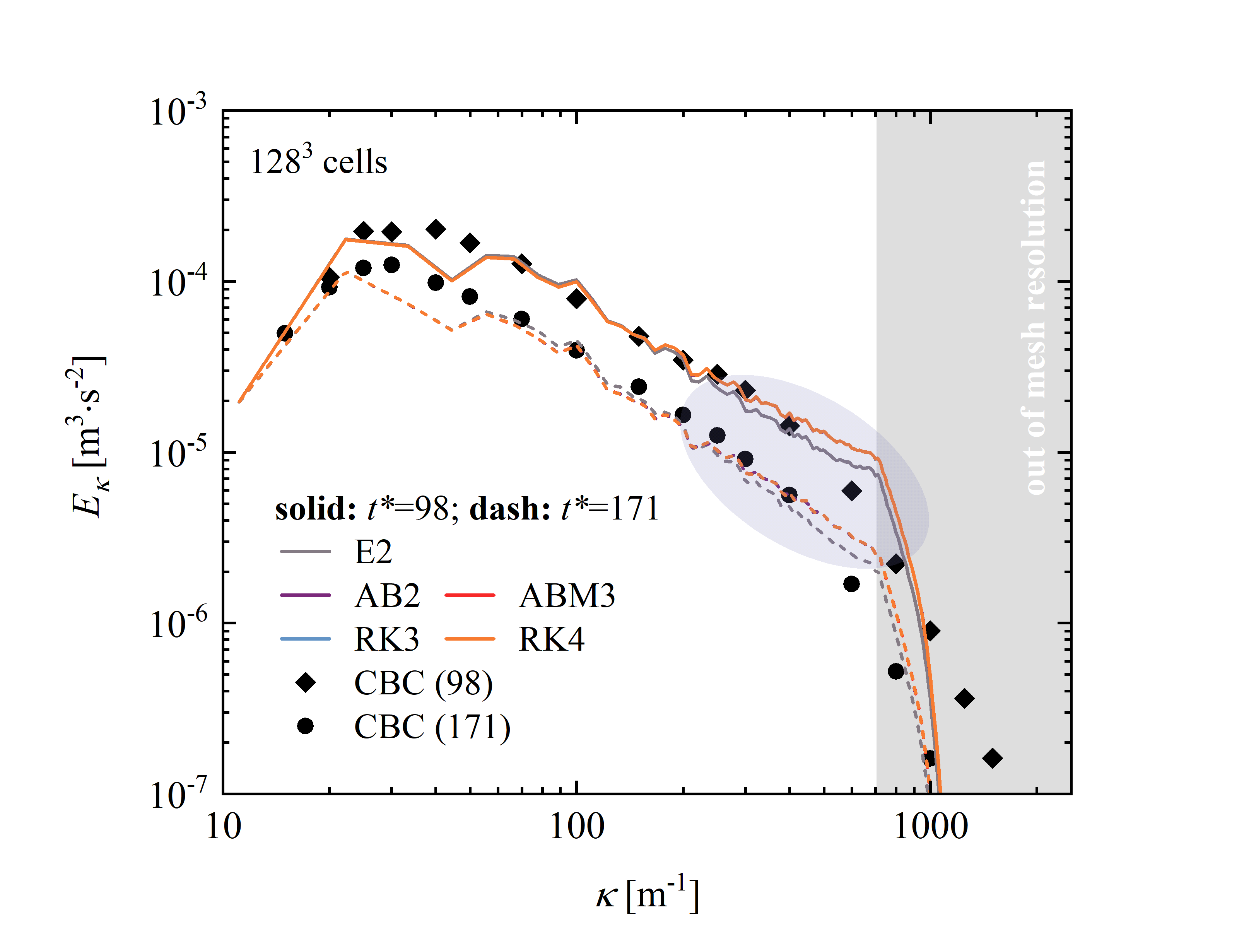}
      \includegraphics[width=8cm]{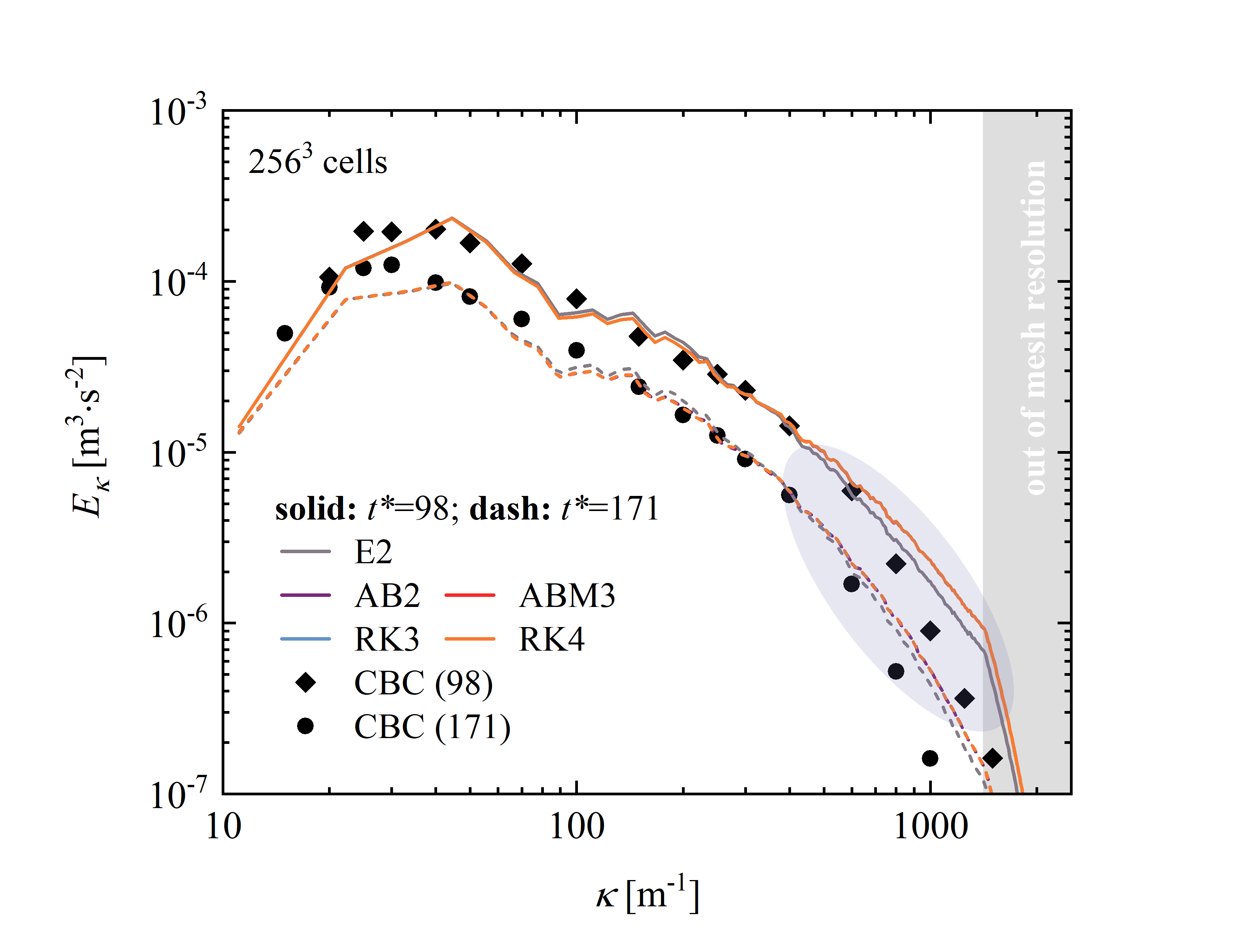}
    }
  }
	\caption{Spectrum distribution of turbulent kinetic energy calculated by different grid resolutions. (\textit{a}) cell number $64^3$; (\textit{b}) cell number $128^3$; (\textit{c}) cell number $256^3$. The reference data are from the literature of \citet{CBC:1971}.}
	\label{fig:DHIT:spectra}
\end{figure}

Fig.\ref{fig:DHIT:spectra} shows the energy spectrum distribution at two moments of DHIT problem, where the solid line and the dashed line represent the data at two different moments respectively. Due to the special case setting, the overall dissipation of the case is insufficient when the number of grids is small: in Fig.\ref{fig:DHIT:spectra}(\textit{a}), there is a clear small-scale energy increase at the 643 grid resolution. However, this phenomenon gradually weakens with the mesh resolution increase. When the inertial scale turbulence is fully resolved using the 2563 grid resolution, the energy spectrum results in Fig.\ref{fig:DHIT:spectra}(\textit{c}) are in good agreement with the experimental data. This is why ILES using the zero-equation sub-grid model often require a high enough grid resolution to correctly capture turbulence.

Under the three grid resolutions, the energy spectra of the four low-dissipative schemes are basically the same, but exhibit apparent difference with backward scheme in the purple region. Among them, the implicit scheme has a small upward shift, while the low-dissipation explicit scheme has a higher one This suggests that the framework's schemes are less dissipative and therefore exhibits greater bias when model dissipation and other numerical dissipation are insufficient equally. Backward scheme has stronger scheme dissipation, so when the overall dissipation is insufficient, part of the numerical dissipation is supplemented, which makes the deviation of the result smaller. The energy spectrum difference caused by the dissipation difference of time discretization decreases gradually with the increase of the number of grids. In the highest resolution grid, the energy spectrum of different formats has been relatively consistent. The results of this case show that in the scale-resolving turbulence simulation, more accurate results do not mean that the dissipation of the scheme is small enough, but are related to the overall numerical method: if a DNS calculation with a high grid accuracy, the results of the low dissipation scheme may be closer to the experimental results than the strong numerical dissipation implicit scheme; However, in an ILES with insufficient grid precision, the results of the low-dissipation scheme may instead have a larger upward shift bias.

\begin{figure}
	\centering
  \fbox{
      \includegraphics[width=8cm]{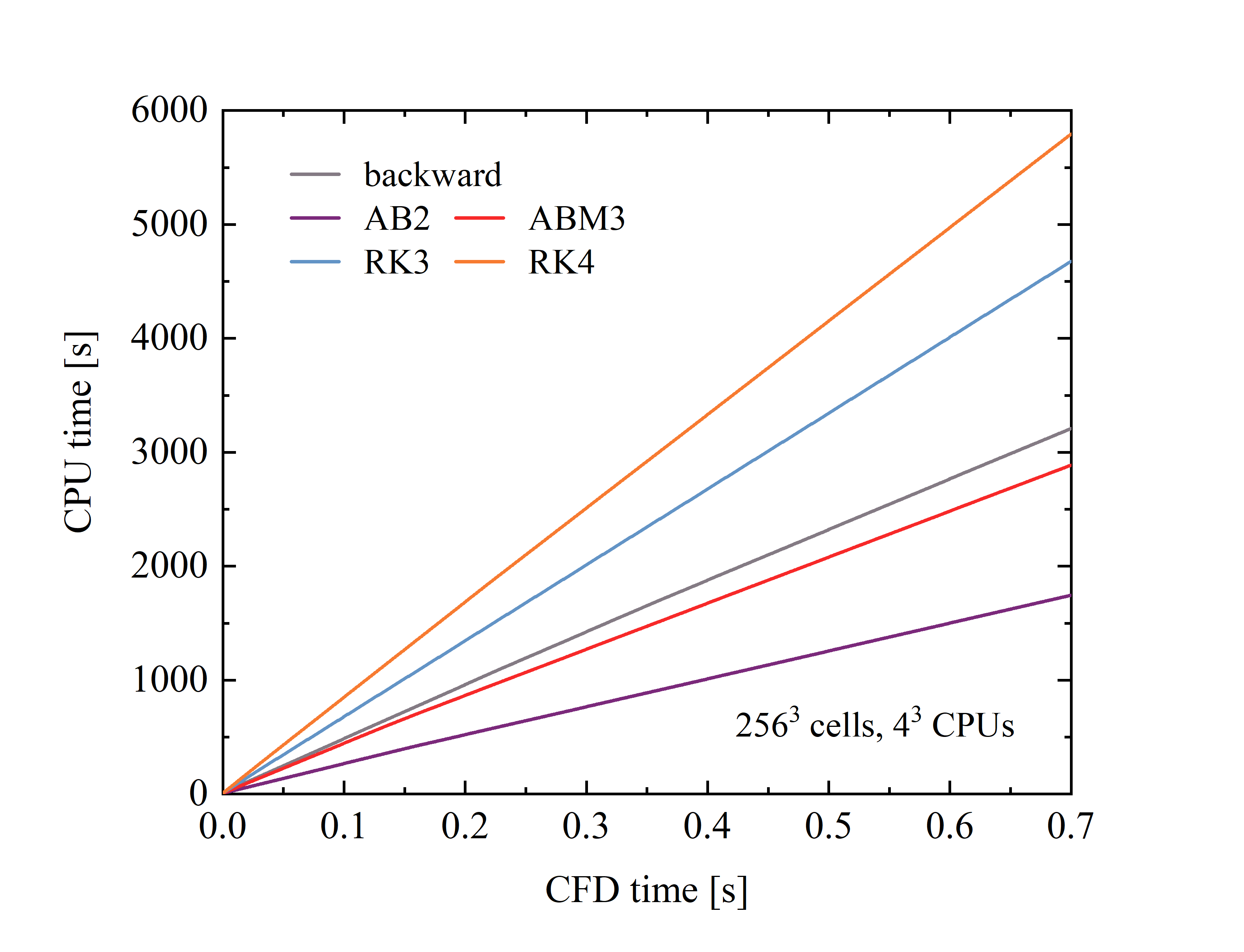}
  }
	\caption{CPU time statistics for different schemes at 2563 mesh resolution.}
	\label{fig:DHIT:speed}
\end{figure}

Fig.\ref{fig:DHIT:speed} shows the total CPU time for DHIT at the $256^3$ grid resolution. In this unsteady problem, the computational speed remains approximately constant throughout the whole process. Compared with RK4, the calculation time of RK3 is significantly shorter, but it still does not reach the expected speed increase. AB2 also has the fastest computation time, while ABM3 has a similar computation speed as backward. The above phenomenon is also consistent with more detailed results in Table \ref{tab:DHIT:speed}. Generally speaking, compared with backward scheme, the linear multi-step method not only has lower scheme dissipation, but also shows advantages in computation speed under the same convergence accuracy in CFD computation.

\begin{table}
  \caption{Computation speed of DHIT at different mesh resolution.}
  \centering
  \begin{tabular}{cccccc}
    \toprule
    \multirow{2}{*}{Mesh} & \multicolumn{5}{c}{Computation speed (steps per second per process)}            \\
    \cmidrule{2-6}
                               & backward    & AB2         & ABM3        & RK3         & RK4         \\
    \midrule
    $64^3$                   & 7.36E-01     & 1.12E+00    & 7.04E-01    & 4.70E-01    & 3.93E-01    \\
    $128^3$                  & 7.09E-02     & 1.18E-01    & 7.24E-02    & 4.75E-02    & 3.82E-02    \\
    $256^3$                  & 6.81E-03     & 1.25E-02    & 7.57E-03    & 4.68E-03    & 3.78E-03    \\
    \bottomrule
    \end{tabular}
  \label{tab:DHIT:speed}
\end{table}

\subsection{Case 4: channel flow}

\begin{figure}
	\centering
  \fbox{
    \parbox[b]{16cm}{
      \centering
      \includegraphics[width=7.9cm]{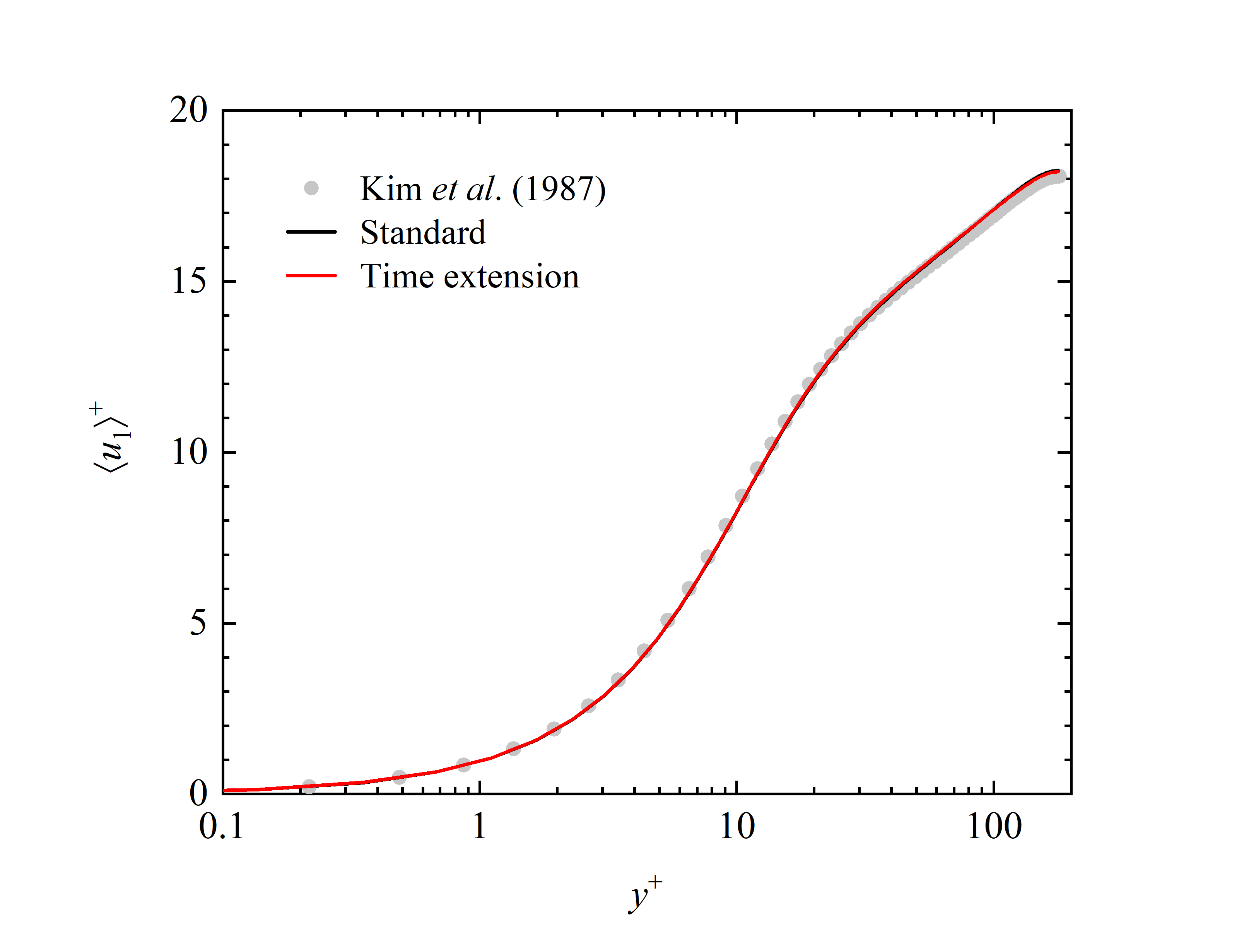}
      \includegraphics[width=7.9cm]{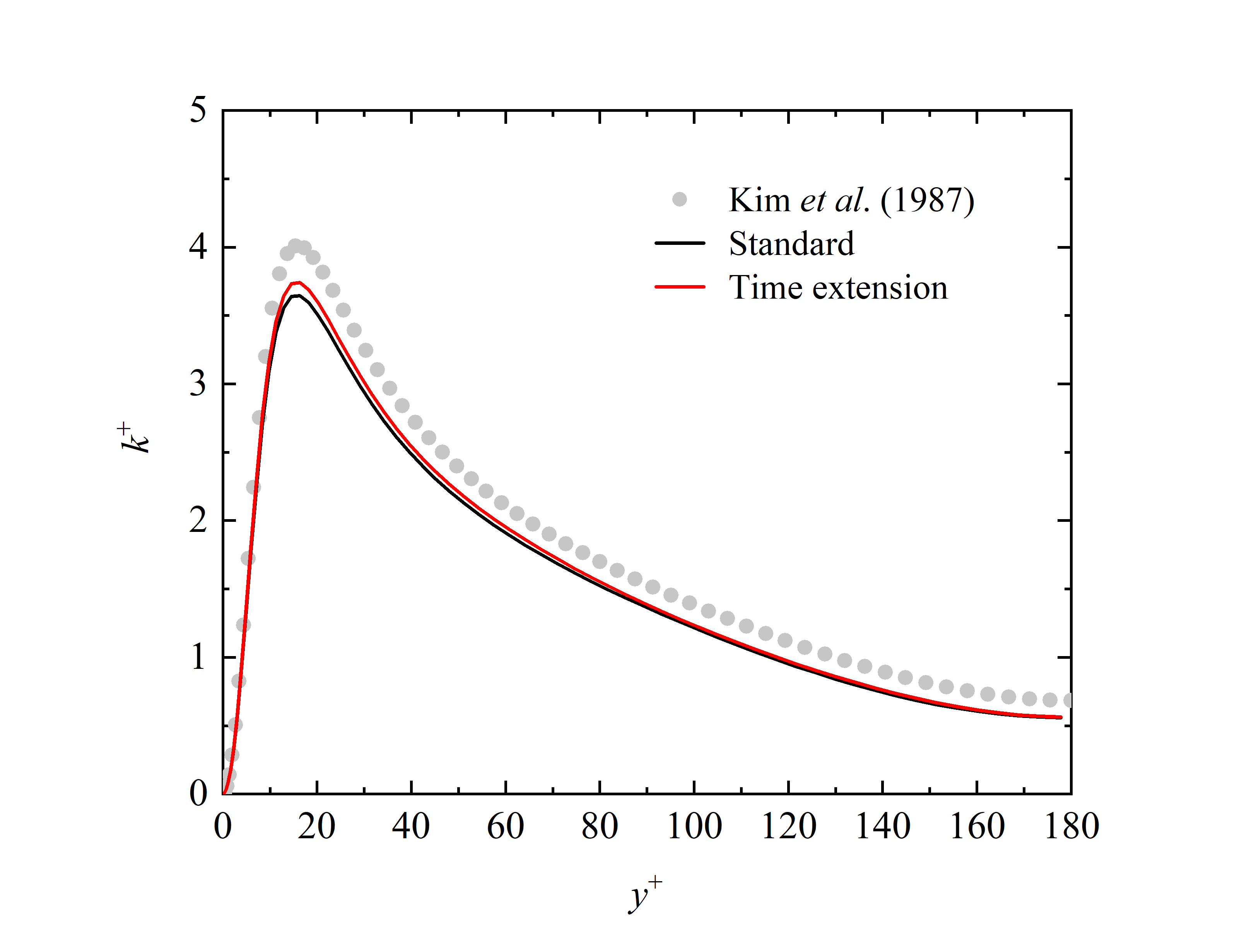}
      \includegraphics[width=7.9cm]{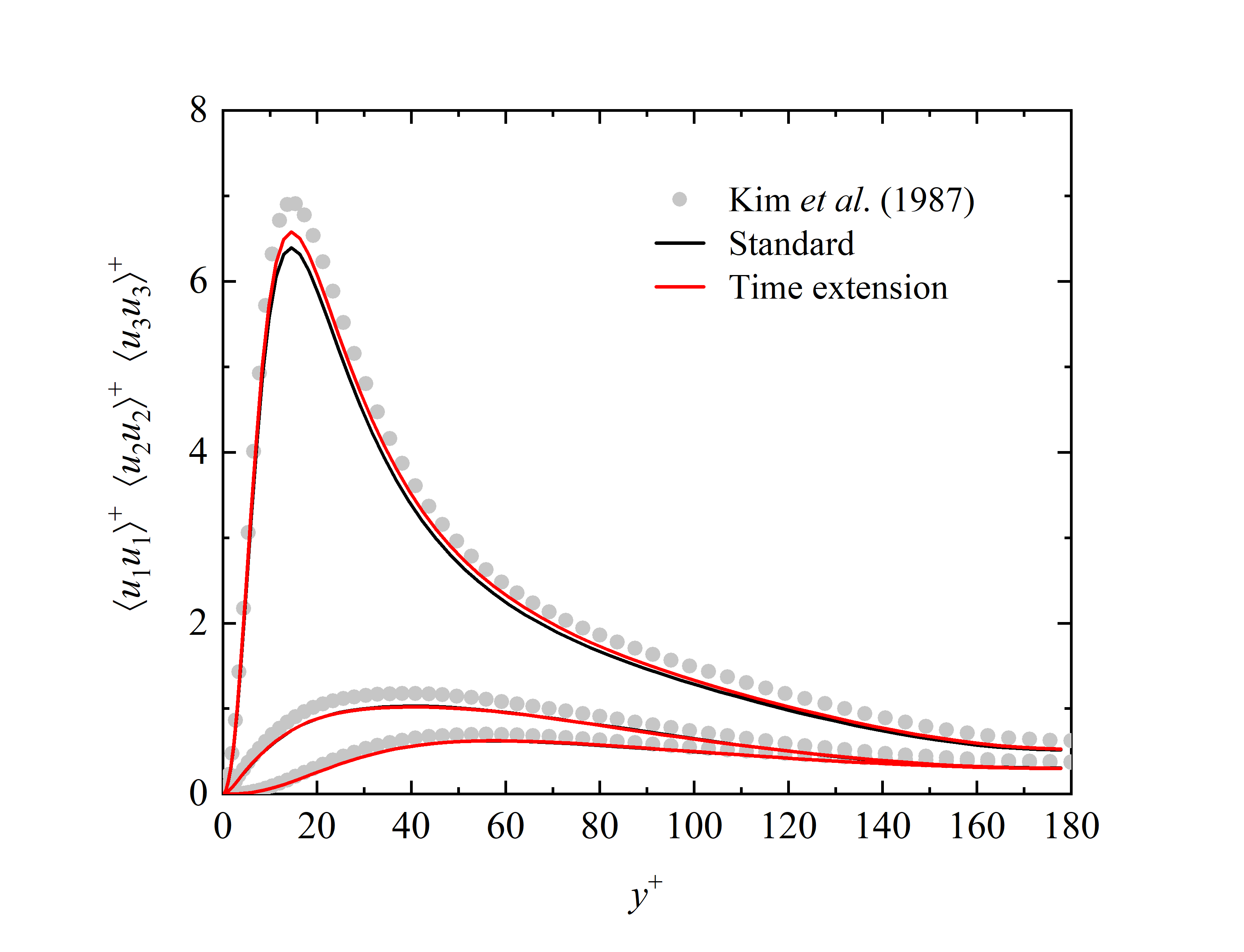}
      \includegraphics[width=7.9cm]{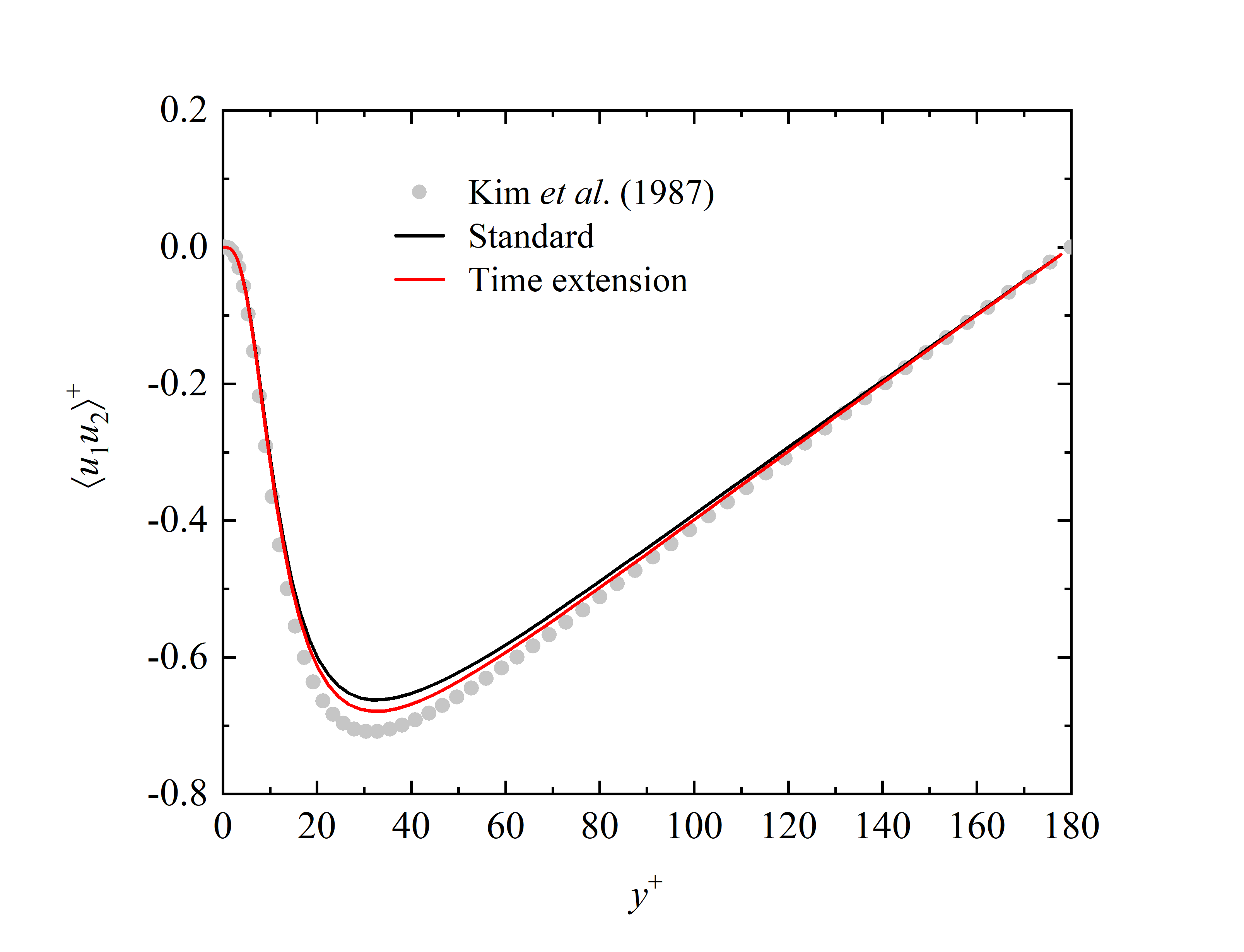}
    }
  }
	\caption{The influence of computation time on the results of implicit time discretization solution. (\textit{a}) velocity boundary layer $\langle u_1 \rangle^+$; (\textit{b}) turbulent kinetic energy $k^+$; (\textit{c}) Reynolds normal stress $\langle u_{i}u_{(i)} \rangle^+$; (\textit{d}) Reynolds shear stress $\langle u_{1}u_{2} \rangle^+$. The reference data are from the literature of \citet{Kim:1987}.}
	\label{fig:PCF:implicit}
\end{figure}

Fig.\ref{fig:PCF:implicit} shows the distribution of the typical quantities of each turbulence calculated by the standard baseline case and the time-extended case. As can be seen from Fig.\ref{fig:PCF:implicit}(\textit{a}), the velocity boundary layer obtained by the solution has a high precision, and the improvement by extending the calculation time is small. The difference between the calculated results and the reference values mainly lies in in the correlation functions in Fig.\ref{fig:PCF:implicit}(\textit{b}), (\textit{c}) and (\textit{d}). Among them, there is a certain error in the shear stress $\langle u_{1}u_{2} \rangle^+$ calculated by backward scheme, which can be reduced by increasing the calculation time. There is also some deviation in the peak region of the streamwise normal stress $\langle u_{1}u_{1} \rangle^+$, which can also be reduced by extending the time of transition calculation and statistical calculation. However, in the calculation results of backward scheme, the system error of wall-normal direction $\langle u_{2}u_{2} \rangle^+$ and spanwise direction $\langle u_{3}u_{3} \rangle^+$ is relatively obvious, and this error does not exhibit apparent improvement when increasing calculation time. Therefore, it seems impossible to improve the accuracy of the results by only increasing the transition and statistical calculation time of the implicit method. Through the test, it was also found that further reducing the step size had no obvious effect on the improvement of $\langle u_{2}u_{2} \rangle^+$ and $\langle u_{3}u_{3} \rangle^+$ accuracy.

\begin{figure}
	\centering
  \fbox{
    \parbox[b]{16cm}{
      \centering
      \includegraphics[width=7.9cm]{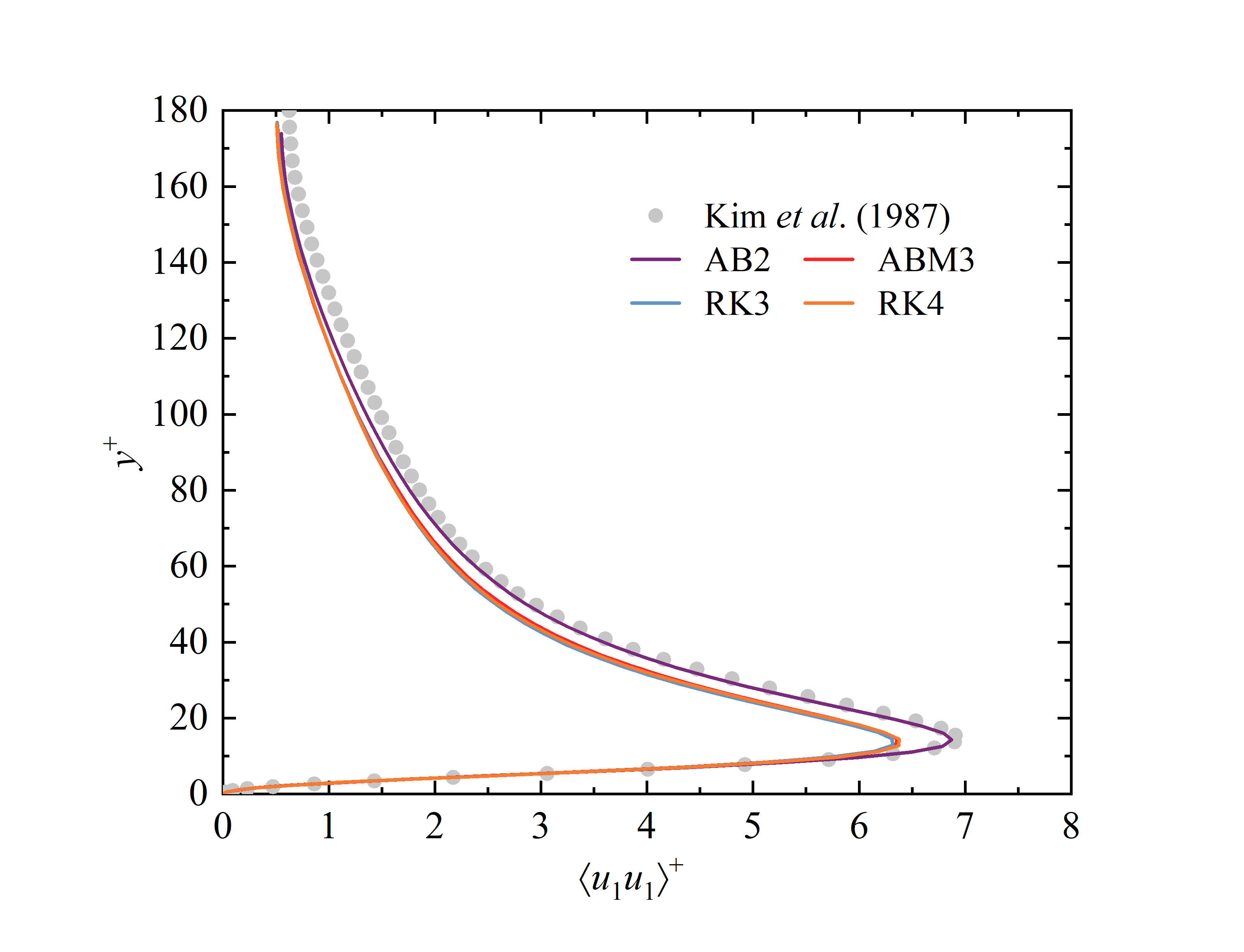}
      \includegraphics[width=7.9cm]{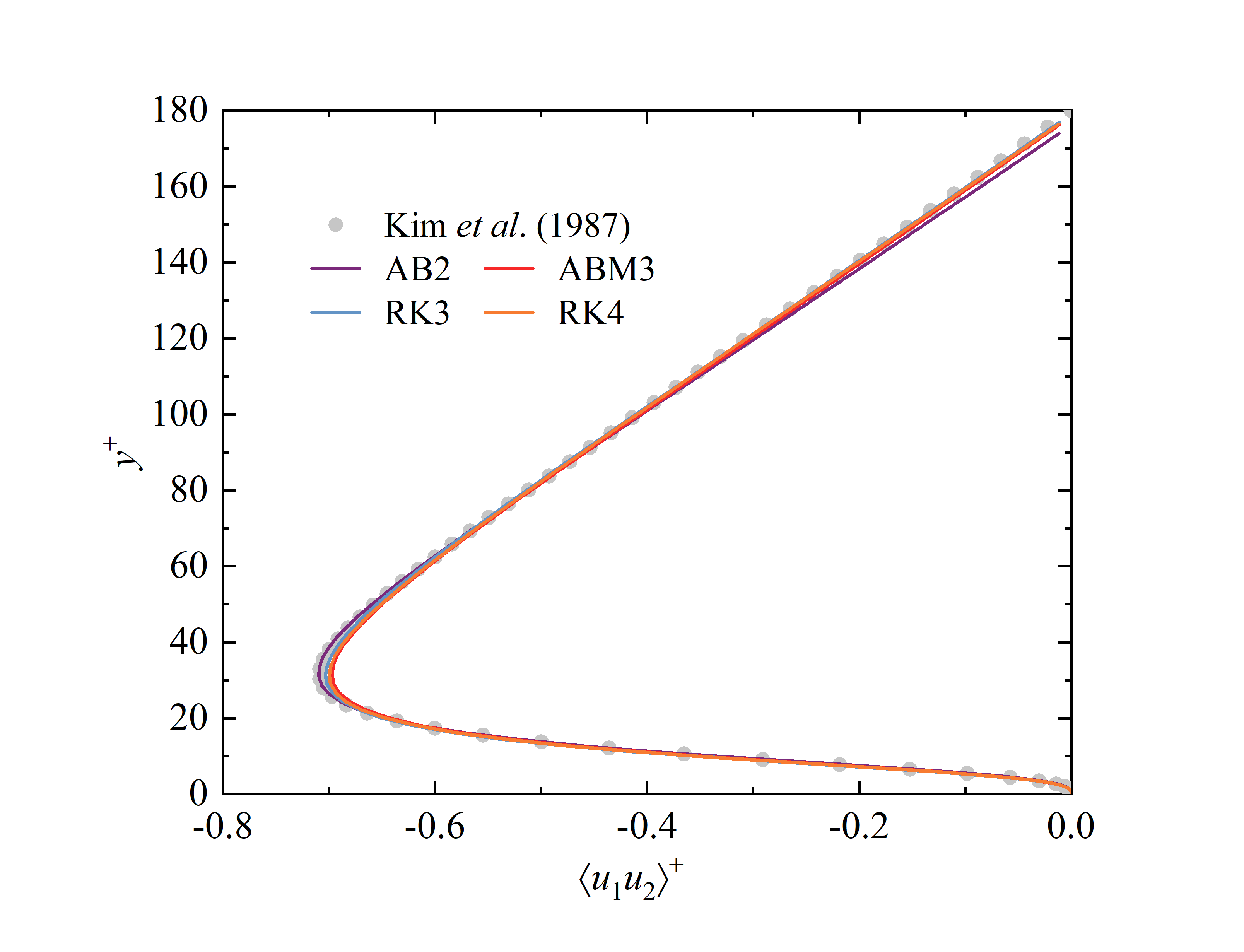}
      \includegraphics[width=7.9cm]{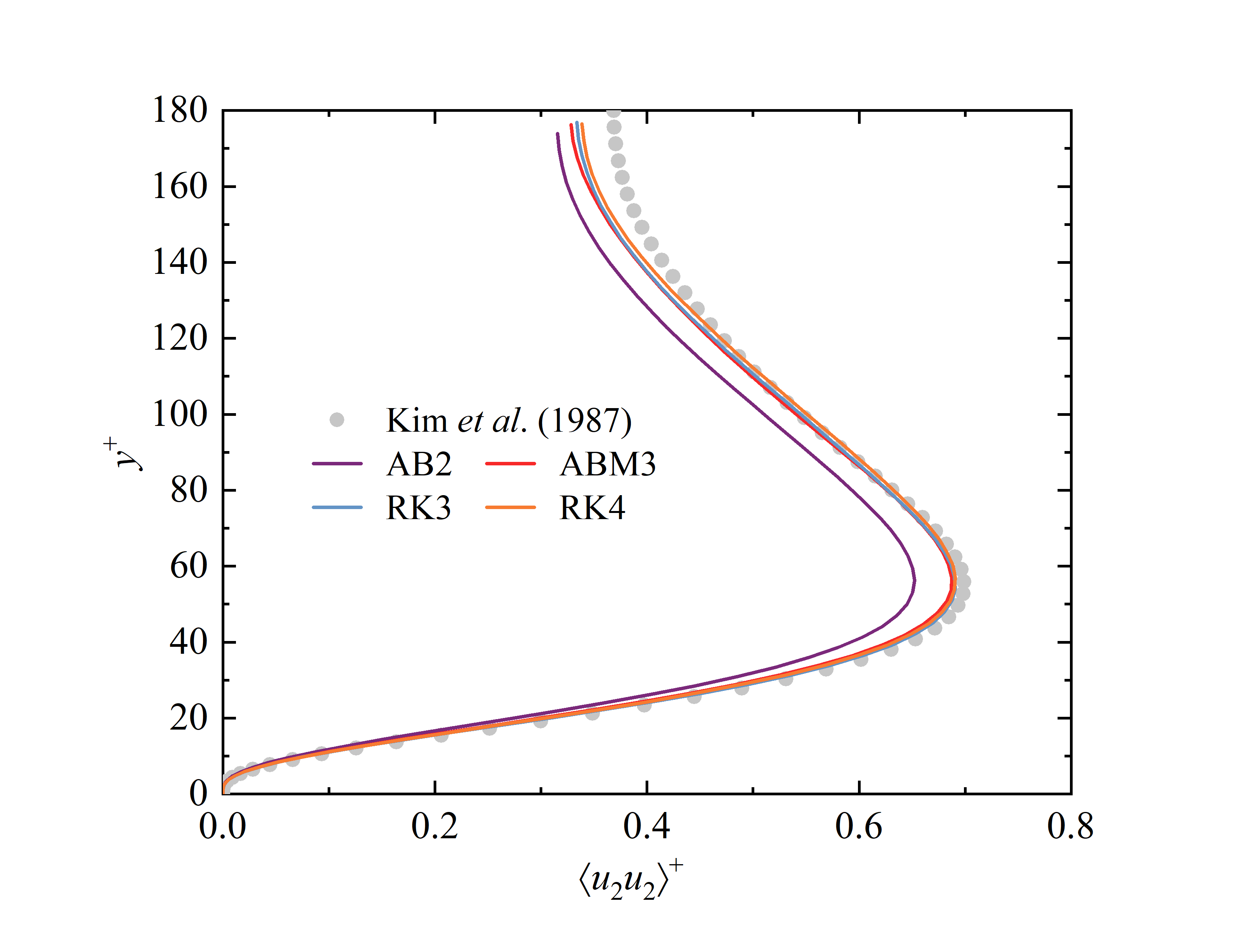}
      \includegraphics[width=7.9cm]{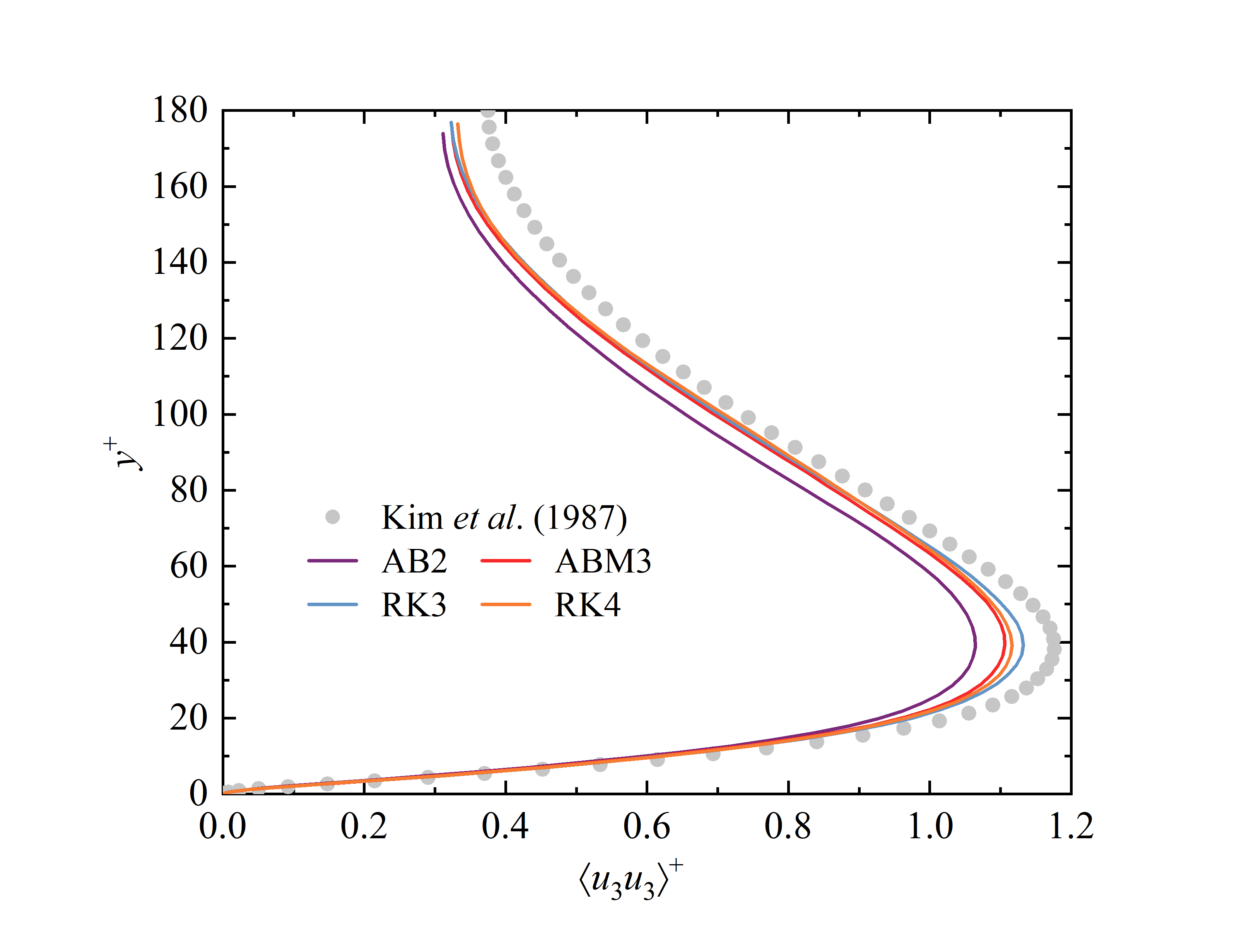}
    }
  }
	\caption{Turbulence correlation calculated by different low-dissipative schemes. (\textit{a}) $\langle u_{1}u_{1} \rangle^+$; (\textit{b}) $\langle u_{1}u_{2} \rangle^+$; (\textit{c}) $\langle u_{2}u_{2} \rangle^+$; (\textit{d}) $\langle u_{3}u_{3} \rangle^+$. The reference data are from the literature of \citet{Kim:1987}.}
	\label{fig:PCF:explicit}
\end{figure}

The turbulence correlation functions calculated by each low-dissipation scheme are shown in Fig.\ref{fig:PCF:explicit}. It is clearly shown in Fig.\ref{fig:PCF:explicit}(\textit{b}) that the accuracy of shear stress $\langle u_{1}u_{2} \rangle^+$ of the four low-dissipation schemes is obviously superior to that of backward scheme. There are slight differences between the distributions of $\langle u_{1}u_{2} \rangle^+$ in different schemes, but they are all fairly accurate However, AB2 with different time step intervals is significantly different from the other three schemes in the normal stress components. In the calculation of the streamwise normal stress $\langle u_{1}u_{1} \rangle^+$, AB2 shows better accuracy than other schemes in the region near the peak (Fig.\ref{fig:PCF:explicit}(\textit{a})). This is due to the higher time resolution used in AB2, so the statistical average calculation can use more time samples for the same total time length, thus accelerating statistical convergence. Therefore, this advantage of AB2 gradually decreases in mainstream regions where turbulence intensity is weaker. Similar to backward scheme in Fig.\ref{fig:PCF:implicit}, if the time of transition calculation and statistical calculation is increased, the accuracy of the other three formats in predicting the streamwise normal stress $\langle u_{1}u_{1} \rangle^+$ can be improved. For the backward scheme systematic error in $\langle u_{2}u_{2} \rangle^+$ and $\langle u_{3}u_{3} \rangle^+$, the calculated results of AB2 in Fig.\ref{fig:PCF:explicit}(\textit{c}) and Fig.\ref{fig:PCF:explicit}(\textit{d}) also have clear error. At this time, the other three schemes with higher order accuracy have significant accuracy improvements. Compared with backward and AB2, the accuracy improvement of the three higher-order method ABM3, RK3 and RK4 is not limited to the peak region, but also has the corresponding accuracy improvement effect in the mainstream region with weaker turbulence. This suggests that the improvement in the higher-order method results is not due to the statistical convergence. In Fig.\ref{fig:PCF:explicit}(\textit{d}), ABM3, RK3, and RK4 also differ slightly, and RK4 with the highest order is not the closest to the reference value. This shows that the influence of errors from other sources besides the discretization order is more influential in this quasi-DNS calculation case when the order or accuracy is high enough, and further increasing the convergence order cannot bring significant improvement.

Although Fig.\ref{fig:PCF:implicit} and Fig.\ref{fig:PCF:explicit} show the advantages of low-dissipation scheme in accuracy compared with backward scheme, and better turbulent time resolution can often be achieved by reducing time step size in order to meet the stability requirements, it is not advisable to blindly pursue the strategy of higher order time discrete scheme. This is because the additional computational cost may result in an order of magnitude change in the computational complexity of the case. Table \ref{tab:PCF:speed} lists the amplification of the total CPU time of the above five cases (four scheme cases + one time extension case) compared with the baseline. It can be seen from the table that although the single step calculation time of the AB2 scheme is faster, the total time required is almost 20 times that of the baseline case due to the large increase in the total time step. Similarly, the two higher-order single-step schemes approximate 24 and 28 times longer. This computational time increase means the difference between the simulation being completed in days or months. In many cases, the accuracy improvement brought by such a large cost is not worthy. A more efficient alternative would be to extend the calculation time, as shown in the table, which requires a time amplification of only 2 to 5 times. In addition, if it is necessary to use a low-dissipative high-order scheme to obtain some higher-precision solutions, it is better to choose a multi-step method with an appropriate number of orders. Because at this time, the accuracy change brought by further increasing the order will be swallowed by the error from other sources, and the fast sped of the multi-step method itself makes the increase of the calculation time be suppressed to a certain extent. For example, the calculation time of ABM3 in Table 5 is 13 times, which has obvious advantages compared with other low-dissipation methods, and the calculation results in Fig.\ref{fig:PCF:explicit} also confirm that it has relatively good accuracy.

\begin{table}
  \caption{The amplification of computation time by different low-dissipation time-discretization schemes.}
  \centering
  \begin{tabular}{ccccccc}
    \toprule
    \multirow{2}{*}{Calculation period} &          & \multicolumn{5}{c}{Time amplification}                   \\
    \cmidrule{2-7}
                                        & backward & Time-expand-backward & AB2    & ABM3   & RK3    & RK4    \\
    \midrule
    Transient                           & 1        & 5.0973               & 20.008 & 13.400 & 24.285 & 28.736 \\
    Statistical                         & 1        & 2.5303               & 19.824 & 13.040 & 23.414 & 27.034 \\
    \bottomrule
    \end{tabular}
  \label{tab:PCF:speed}
\end{table}

\section{Conclusions}
\label{sec:conclusions}

In this paper, we establish a low-dissipation projection solver framework based on the open-source C++ libraries, which integrates a variety of high order time discretization schemes, including linear single-step method and linear multi-step method. Among them, in addition to the common Runge--Kutta family of single-step schemes, this study implement two types of multi-step schemes: Adams--Bashforth method and Adams--Bashforth--Moutton method. The unified solver obtained by this framework can dynamically select the specified low-dissipative scheme in the calculation. Through a series of test cases carried out on five schemes including second-order implicit backward Euler scheme, the solution accuracy and computation speed of low-dissipative universal framework and different low-dissipation schemes are comprehensively evaluated. The main conclusions are as follows.

(1) A variety of low-dissipation high-order schemes based on the projection framework have reached the theoretical convergence order in practical verification, and lower scheme dissipation is shown in scale-resolving turbulence simulation. The test cases based on high resolution turbulence simulation show that the four low-dissipative schemes all have better performance in forecasting accuracy than backward scheme, but there are differences between them due to different types and orders of methods. All the test cases show effectiveness and usefulness of this general framework in time discretization studies

(2) Compared with the linear single-step scheme of the same order, the linear multi-step scheme has a very obvious advantage in computational efficiency. In some cases, the computation time of the multi-step scheme with higher order is even lower than that of the second-order implicit scheme. The main cost of the multi-step scheme comes from the increased memory requirements, which can be fully met in today's conventional high-performance computers.

(3) Computational stability is the main problem in the application of explicit low-dissipation schemes in CFD. In order to ensure that the calculation does not diverge, it is sometimes necessary to drastically reduce the time step. From the point of view of ensuring the accuracy of the result and sufficiently low numerical dissipation, the reduction of the time step is advantageous; But the potential harm is that too small a time step and the amount of computation due to multiple stages of high-order scheme itself can lead to an order of magnitude increase in computation time.

(4) When the stability problem occurs, considering the solution accuracy versus computation speed, it is found that a multi-step scheme with moderate order may be the best choice. One practical difficulty is that it is often difficult for researchers to determine which high-order scheme is optimal before performing CFD calculations. In this case, using the low-dissipation solution framework to dynamically select a variety of different schemes or embed the researcher's developing schemes for test becomes a suitable and attractive choice.

\section*{Code availability}

The source code to reproduce the results in this paper will be openly available on GitHub at \href{https://github.com/Fracturist/High-Order-Time-Scheme-Framework}{High-Order-Time-Scheme-Framework} upon publication.

\bibliographystyle{unsrtnat}
\bibliography{lowDissFrame}

\end{document}